\documentclass[aps,prx,groupedaddress,showpacs,twocolumn,longbibliography]{revtex4-1}
\usepackage[utf8]{inputenc}
\usepackage{hyperref}
\usepackage{graphicx}  
\usepackage{dcolumn}   
\usepackage{bm}        
\usepackage{amssymb,amsmath}   
\usepackage{uri}

\usepackage[x11names, rgb, html]{xcolor}
\definecolor{sbase03}{HTML}{002B36}
\definecolor{sbase02}{HTML}{073642}
\definecolor{sbase01}{HTML}{586E75}
\definecolor{sbase00}{HTML}{657B83}
\definecolor{sbase0}{HTML}{839496}
\definecolor{sbase1}{HTML}{93A1A1}
\definecolor{sbase2}{HTML}{EEE8D5}
\definecolor{sbase3}{HTML}{FDF6E3}
\definecolor{syellow}{HTML}{B58900}
\definecolor{sorange}{HTML}{CB4B16}
\definecolor{sred}{HTML}{DC322F}
\definecolor{smagenta}{HTML}{D33682}
\definecolor{sviolet}{HTML}{6C71C4}
\definecolor{sblue}{HTML}{268BD2}
\definecolor{scyan}{HTML}{2AA198}
\definecolor{sgreen}{HTML}{859900}
\hypersetup{
  colorlinks = true,
  allcolors = {blue}
}
\usepackage{color}

\DeclareMathOperator{\erf}{erf}
\DeclareMathOperator{\sign}{sign}

\begin{document}
\title{Controlling particle currents with evaporation and 
resetting from an interval}
\author{Gennaro Tucci$^{1}$}
\author{Andrea Gambassi$^{1}$}
\author{Shamik Gupta$^{2,3}$}
\author{\'Edgar Rold\'an$^{3}$}\email{edgar@ictp.it}
\affiliation{$^1$SISSA -- International School for Advanced Studies and INFN, via Bonomea 265, 34136 Trieste, Italy \\
${^2}$Department of Physics, Ramakrishna Mission Vivekananda Educational and
Research Institute, Belur Math, Howrah 711202, India\\
${^3}$ICTP - The Abdus Salam International Centre for Theoretical Physics, Strada Costiera 11, 34151, Trieste, Italy}

\begin{abstract} 
 
We investigate the Brownian diffusion of particles in one spatial dimension and in the presence of finite regions within which particles can either evaporate or be reset to a given location.  For open boundary conditions, we highlight the appearance of a Brownian yet non-Gaussian diffusion: at long times, the particle distribution is non-Gaussian but its variance grows linearly in time. Moreover, we show that the effective diffusion coefficient of the particles in such systems is bounded from below by $(1-2/\pi)$ times their bare diffusion coefficient. For periodic boundary conditions, i.e., for diffusion on a ring with resetting, we demonstrate a ``gauge invariance'' of the spatial particle distribution for different choices of the resetting probability currents, in both stationary and non-stationary regimes. Finally, we apply our findings to a stochastic biophysical model for the motion of  RNA polymerases during transcriptional pauses, deriving analytically the distribution of the length of cleaved RNA transcripts and the efficiency of RNA cleavage in backtrack recovery. 

   \end{abstract} 
\maketitle

\section{Introduction}
Stochastic processes whereby incremental changes are interspersed with
sudden and large changes occurring at unpredictable times are common in
nature~\cite{odde1995kinetics,julicher1997modeling,brugues2012nucleation,pavin2014swinging,evans2011diffusion,kusmierz2015optimal,luca,durang,gupta2014fluctuating,harris2017phase,lisica2016mechanisms,mora2015physical,besga2020optimal,tal2020experimental}. Examples range from epidemics (e.g., Covid-19), to financial
market (e.g., the 2008 crisis) and biology (e.g., the catastrophic events of sudden
shrinkage during the polymerization of a microtubule \cite{odde1995kinetics,brugues2012nucleation,pavin2014swinging}, the flashing ratchet mechanism of molecular motors~\cite{julicher1997modeling}, etc.). Recent
development of  the theoretical  ``stochastic resetting" framework aims
to describe features of such stochastic processes. 
A paradigmatic model for these phenomena is provided by a
Brownian particle diffusing and also resetting instantaneously its position to a fixed value at exponentially-distributed random time intervals~\cite{evans2011diffusion}. 
 The concept of stochastic resetting has been invoked in many different fields
 such as first-passage properties~\cite{kusmierz2015optimal,pal1},  continuous-time random walk~\cite{PhysRevE.93.022106}, foraging~\cite{luca,boyer2014random,bhat2016stochastic,ray2019peclet}, reaction-diffusion
 models~\cite{durang}, fluctuating
 interfaces~\cite{gupta2014fluctuating,nagar2016}, exclusion
 processes~\cite{urna},  phase
 transitions~\cite{harris2017phase},  large deviations~\cite{coghi2020large}, RNA
 transcription~\cite{lisica2016mechanisms,roldan2016stochastic}, quantum
 dynamics~\cite{krishnendu}, cellular sensing~\cite{mora2015physical}, population dynamics~\cite{garcia2019linking}, stochastic
 thermodynamics~\cite{udo-reset}, and active matter~\cite{maes2017induced}, see Ref.~\cite{majumdar-review} for a recent review.

One of the main physical consequences  of resetting  is its ability to induce a
stationary state in systems that otherwise would not allow such a state to exist. A paradigmatic example is the Brownian motion, where the addition of spatially homogeneous resetting, in any spatial dimensionality $d$ \cite{evans2011diffusion,Evans_2014}, ensures a stationary state even in the absence of confining boundary conditions and/or potentials.
In the recent past, a variety of situations have been studied within
this framework, e.g., Brownian particles resetting to a generic spatial distribution, under the action of an external potential~\cite{path_integral}, and for various choices of the resetting time probability distribution~\cite{Shamik}. Most of these studies consider resetting phenomena with open boundary conditions. 
However, excepting for rare instances such as Refs. \cite{circle,Christou_2015}, not much is known about systems that experience
stochastic resetting in a closed geometry, such as periodic boundary
conditions in one dimension. In particular, how do topological constraints
in $d=1$ (e.g., boundary conditions) affect the stationary and dynamical properties of diffusions with stochastic resetting?
 
In this work, we study minimal stochastic models of Brownian particles evaporating or resetting 
from a finite one-dimensional region  with either open, periodic and/or absorbing boundary
conditions. As the first case, we consider the \textit{tunnelling} of Brownian particles across an interval  with a constant evaporation rate. The solution of such \textit{Brownian tunnelling} model  allows us to tackle  the solution  for the problem with resetting~\cite{path_integral}. In the context stochastic resetting, we interpret evaporation as a resetting to an absorbing point. In the case of resetting with open boundary conditions, we highlight the 
appearance of   \textit{Brownian yet non-Gaussian} diffusion~\cite{chubynsky2014diffusing,chechkin2017brownian,sposini2018random,miotto2019length} at
large times, meaning that although the distribution of the particle
position is non-Gaussian, its mean-squared displacement grows linearly in
time at leading order.

 Our key findings concern the transport properties of
diffusion processes with resetting and periodic boundary
conditions. 
For these systems, a stationary state exists and 
we can define a probability current with non-vanishing stationary value. We reveal that the actual value of this current depends genuinely on the local \textit{direction} of resetting. For example, particles on a ring can reset in clockwise or
counterclockwise direction and although this results in the
\textit{same} (stationary) distribution, it leads to a drastic
change in the behavior of the (stationary) current. 
Accordingly, we show that currents provide useful and insightful information about
the underlying non-equilibrium resetting process, in addition to that conveyed by
the probability distribution.
Drawing an analogy with field theory, we
dub this result ``resetting gauge invariance": a whole class of
dynamics, all with very different behavior of the current, results in the same distribution, both in stationary and non-stationary regimes.

We illustrate our findings  by
analyzing several prototypical stochastic models, using as building
blocks exact results that we derive by extending the path-integral formalism
for resetting~\cite{path_integral}, first for models of diffusion with resetting or
evaporation occurring with a fixed rate only within a small spatial
region. We also show that our approach, besides being of interest for physics, applies usefully to a stochastic model of the motion of a biological system, i.e., RNA polymerases during transcriptional pauses, shedding light on  available experimental results~\cite{lisica2016mechanisms,roldan2016stochastic}. 

The rest of the paper is organized as follows. In Sec.~\ref{Sec:2}, we address the problem of the Brownian particle in $d=1$ controlled homogeneously in space on a finite interval by resetting or evaporation. In 
Sec.~\ref{Sec:3}, we study the average and variance of the particle dynamics described in \ref{Sec:2}, focusing on their behavior at long times. In Sec.~\ref{Sec:4}, we discuss  particle currents due to diffusion and resetting  in one-dimensional systems with  periodic boundary conditions. In Sec.~\ref{Sec:5}, we apply our results to the fluctuating motion of RNA polymerases along a DNA template during transcriptional pauses. We conclude the paper with a discussion in Sec.~\ref{Sec:6}. We provide  the exact analytical calculations and  further numerical results in the Appendices.

\section{Controlling diffusion}\label{Sec:2}

We consider a single Brownian particle
in $d=1$, with diffusion coefficient~$D$ and initial
position $x(t=0)=x_0$. Within the time interval $[t,t+\text{d}t]$ with
$t\geq 0$, the particle at position $x(t)$ can either diffuse, or, be controlled by an external agent with a
probability $r_c(x(t))\text{d}t$ that depends on its current location. We consider two types of control mechanisms, labeled by the
parameter $\sigma_c$: (i) evaporation ($\sigma_c=0$), which annihilates the particle; and (ii) resetting ($\sigma_c=1$), which moves instantaneously the particle to a prescribed position $x_r$
that may be different from $x_0$. The conditional
probability density $P_t(x)\equiv P(x,t|x_0,0)$ of the particle position $x$ at time $t$ obeys a Fokker-Planck
equation with source terms:
\begin{equation}\label{ME}
\begin{aligned}
\partial_t P_t(x) -D \partial^2_x  P_t(x)=-r_c(x) P_t(x)+\sigma_c R_t\,\delta(x-x_r),
\end{aligned}
\end{equation}
where $R_t\equiv \int\mathrm{d}y \,r_c(y) P_t(y)$ is the fraction of
particles that are controlled at time $t$. In Eq.~\eqref{ME}, the second
term on the left-hand side accounts for diffusion of the particle and
the source terms on the right-hand side account for the probability loss
due to control and the probability gain when resetting is applied. In
this work, we consider a control exerted on the particle at a rate $r$ only when it is located within the interval $[-a,a]$, i.e.,
\begin{equation}\label{rx}
r_c(x)=r\, \theta (a-|x|),
\end{equation}
 where 
$\theta$ is the Heaviside step function. In words, the control is exerted on the particle at a rate
 $r$ only when the particle is located within the interval $[-a,a]$. This choice of the resetting profile turns out to be more realistic for experimental realizations~\cite{tal2020experimental,besga2020optimal}  with respect to the ideal, spatially homogeneous case, which can be retrieved $a\rightarrow\infty$. We
 will discuss a collection of transport phenomena described by
 Eqs.~\eqref{ME} and \eqref{rx}, together with appropriate boundary
 conditions introduced below.

Deriving exact statistics for the aforementioned class of models by
solving Eq.~\eqref{ME} is a difficult task. Here,
we extend and employ the recently-introduced path-integral approach to resetting~\cite{path_integral}, in order to tackle analytical calculations. 
 First, we evaluate the probability $P_{\rm nc}(x,t|x_0,0)\text{d}x$  to
 find at time $t$ a particle in $[x,x+\text{d}x)$ that has
 experienced no control in the past; $P_{\rm nc}$ satisfies Eq.~(\ref{ME}) with $\sigma_c=0$,
 and, hence, coincides with the probability density of a diffusing
 particle with a space-dependent evaporation rate $r_c(x)$. Furthermore,
 from $P_{\rm nc}$, one can derive the probability density
 for a particle to undergo a control, evaporation or resetting, for the first time at time $t$, as $P_{\rm res}(t|x_0)=\int_{-\infty}^{+\infty}\mathrm{d}y\,r_c(y)\,P_{\rm nc}(y,t|x_0,0)$. 
The general solution of Eq. (\ref{ME}) is given by $P(x,t|x_0,0)= P_{\rm
nc}(x,t|x_0,0)+\sigma_c\int_0^t \mathrm{d}\tau
\int\mathrm{d}y\,r_c(y)P(y,t-\tau|x_0) P_{\rm nc}(x,t|x_r,t-\tau)$~\cite{path_integral},
where the first term on the right-hand side represents the contribution
of particles whose trajectories reach $x$ at time $t$ without undergoing
any control, while  the second  accounts for particles that reset for the
last time at the intermediate time $t-\tau$ and then freely diffuse
starting from $x_r$ (see Appendix \ref{Appendix:P}).  Below, we will use these relations to obtain exact analytical expressions for a variety of cases with resetting and evaporation.

\begin{figure}[t!]
\includegraphics[scale=0.55]{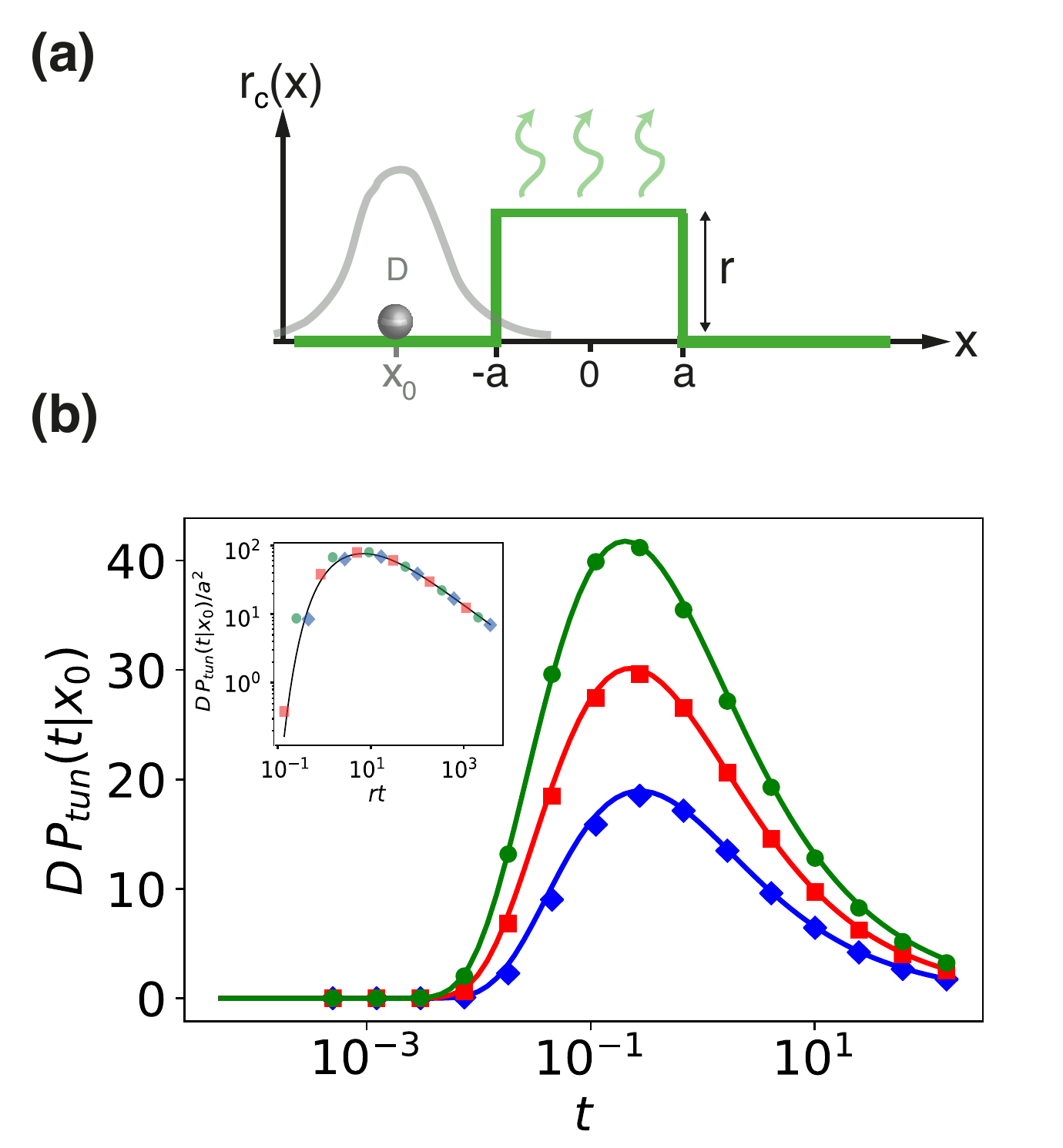}
\caption{\textbf{Diffusion with evaporation within a finite region (Brownian tunnelling)}. \textbf{(a)} Sketch of the model: a Brownian particle (gray sphere) moves in one dimension from an initial position $x_0$ with diffusion coefficient $D$. The particle evaporates (light green arrows) with rate $r$ only from a spatial window $[-a,a]$, i.e., with the evaporation rate $r(x)$ in Eq.~\eqref{rx} (green line). 
\textbf{(b)} Fraction of "tunnelled" particles $P_{\rm tun}(t|x_0)$ (multiplied by $D$) at time $t$ that survived evaporation by reaching the region $x \ge a$: results from numerical simulations (thick lines) and analytical calculations (solid lines). Here $x_0=-3.2$, $r=25$, $D=200$ (green), $150$ (red), $100$ (blue) and $a=0.05\times \sqrt{D}$ respectively. Simulations were done using the Euler integration scheme with time step $\Delta t =5\times 10^{-4}$, and $N=10^4$ runs. The inset shows the collapse of the three curves in dimensionless units.
\label{fig:tunneling}}
\end{figure}

\textit{Brownian tunnelling.} We first consider a minimal
model -- which we call ``Brownian tunnelling" -- given by a Brownian particle moving in $d=1$ and subject to the evaporation rate in Eq.~\eqref{rx}, see Fig.~\ref{fig:tunneling}a. Its Fokker-Planck equation~\eqref{ME}  corresponds to a Schr\"odinger equation $i\hbar \partial_\tau \Psi_\tau(x) = -(\hbar^2/2m)\partial^2_x \Psi_\tau(x) + V(x) \Psi_\tau(x)$,
in:  imaginary time $\tau=-it$, with effective mass $m=\hbar/2D$ and
effective quantum barrier potential
${V(x)=\hbar\,r_c(x)}$~\cite{path_integral}.  Building on the analogy with tunnelling through a quantum barrier, we  compute  the probability ${P_{\rm tun}(t|x_0)\equiv \int_a^\infty \mathrm{d}x P_{\rm nc}(x,t|x_0)}$ for a particle starting at $x_0<-a$ to be found at any point $x>a$ at time $t$, i.e., the probability for a Brownian particle to ``tunnel" through the evaporation window. 
In particular, we derive the analytical expression for  the Laplace transform ${\tilde{P}_{\rm tun}(s|x_0)\equiv \int_0^\infty \mathrm{d}t\exp(-st) P_{\rm
tun}(t|x_0)}$ of the Brownian tunnelling probability:
\begin{equation}\label{eq:tunneling}
\tilde{P}_{\rm tun}(s|x_0)=
\frac{(\nu/D\mu)\exp\left((a+x_0)\mu\right)}{2\mu\nu\cosh\left(2a\nu\right)+(\mu^2+\nu^2)\sinh\left(2a\nu\right)},
\end{equation}
with $\mu\equiv\sqrt{s/D},~\nu\equiv\sqrt{(s+r)/D}$.
Equation~\eqref{eq:tunneling} suggests a natural set of dimensionless
quantities:
\begin{equation}
y\equiv x/a,\;\;y_r\equiv x_r/a,\;\;\tau_r\equiv rt,\;\;\rho\equiv a\sqrt{r/D}.
\label{eq:dimensionless}
\end{equation}
In Fig.~\ref{fig:tunneling}b we show the comparison between numerical simulations for $P_{\rm tun}$ and the numerical inverse Laplace
transform of  Eq.~\eqref{eq:tunneling}.
Upon rescaling variables, the rescaled tunnelling probability $P_{\rm tun} \times D/a^2$ depends only on one parameter~$\rho$ and the variables $\tau_r$ and $y_r$ (Fig.~\ref{fig:tunneling}b inset).
Furthermore, Eq.~\eqref{eq:tunneling} provides information on the long-time
limit and moments of $P_{\rm tun}(t|x_0)$ by considering the series expansion of
$\tilde{P}_{\rm tun}(s|x_0)$ for small $s$, for which we obtain $\tilde{P}_{\rm tun}(s|x_0)=
\left[\sqrt{s\,r}\sinh\left(2\rho\right)\right]^{-1}+O(s^0)$. The presence
of half-integer powers implies that the integer moments of the tunnelling probability diverge, due to the existence of many trajectories that never cross the resetting region.
In fact, for large times and arbitrary $x_0$, $P_{\rm tun}(t|x_0)\sim
t^{-1/2}$. 
\begin{figure}[t!]
\centering
\includegraphics[scale=0.55]{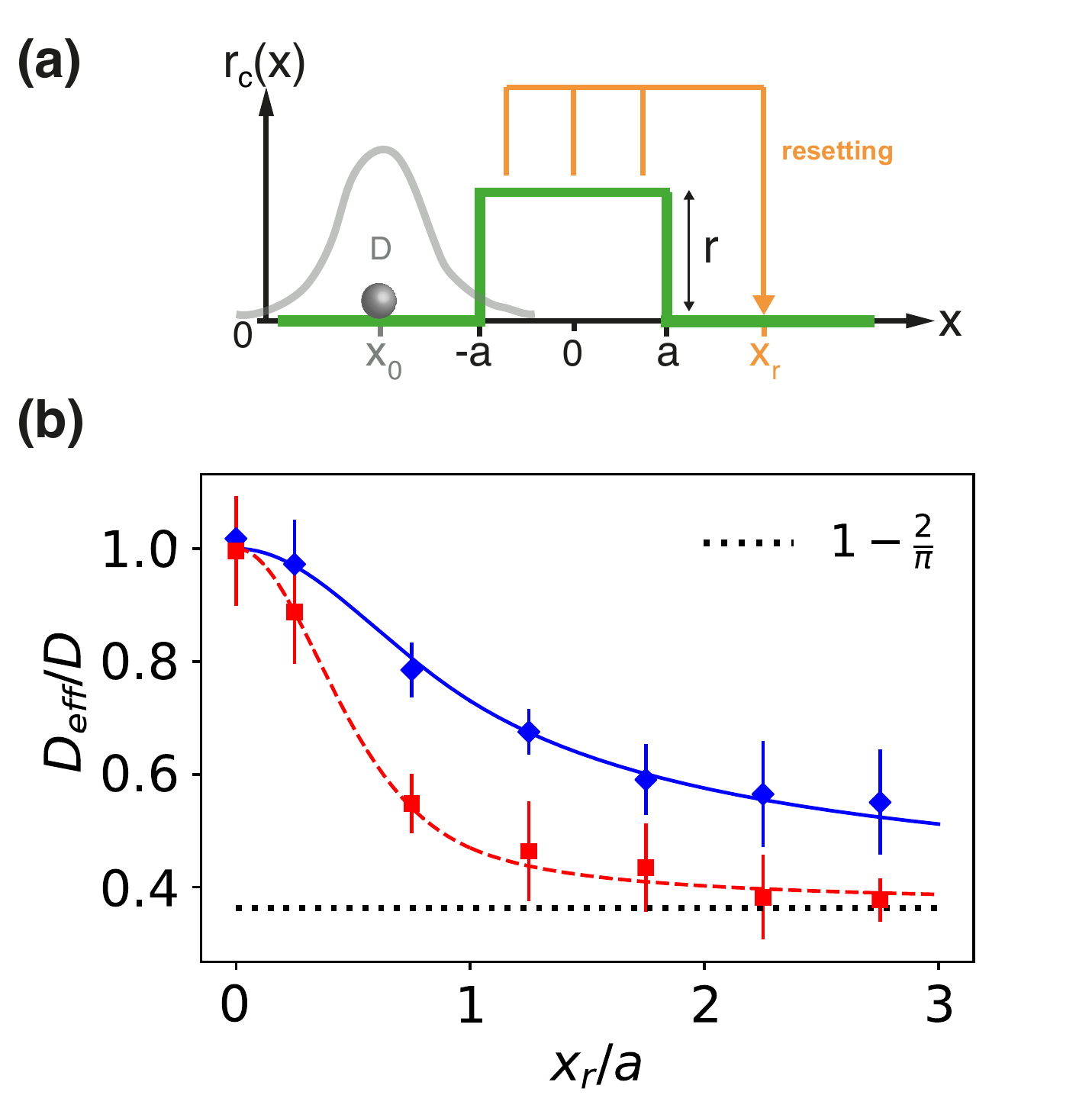}
\caption{\textbf{Diffusion with stochastic resetting from an interval}. (a) Sketch of the model: diffusion of a Brownian particle (gray sphere) with initial position $x_0$, diffusion coefficient $D$, and resetting at rate $r$ from an interval to a resetting destination $x_r$. (b) Effective diffusion coefficient $D_{\rm eff}=\lim_{t\to\infty} \sigma^2(t)/2t$, with $\sigma^2(t)$ the variance of the position, in units of the ''bare'' diffusion coefficient $D$, as a function of normalized resetting point $y_r=x_r/a$. The simulations (symbols) and the analytical expression in Eq.\eqref{eq:defffull} (solid lines) correspond to $\rho\simeq 1.12$ realized with $r=2.5$, $a=5$, $D=50$ (blue diamonds and solid line)  and $\rho\simeq 1.89$ realized with $r=5$, $a=5$, $D=35$ (red squares and dashed line). The horizontal dotted line indicates the theoretical lower bound predicted by Eq.~\eqref{eq:Deff}.  The simulations were done using  Euler's numerical integration scheme with time step $\Delta t =5\times 10^{-2}$, time duration $t=10^3$ and $N=10^5$ runs.}\label{fig:Deff1}
\end{figure}

\section{Minimal diffusivity for a resetting interval}\label{Sec:3}

\begin{figure}
\centering
\includegraphics[scale=0.55]{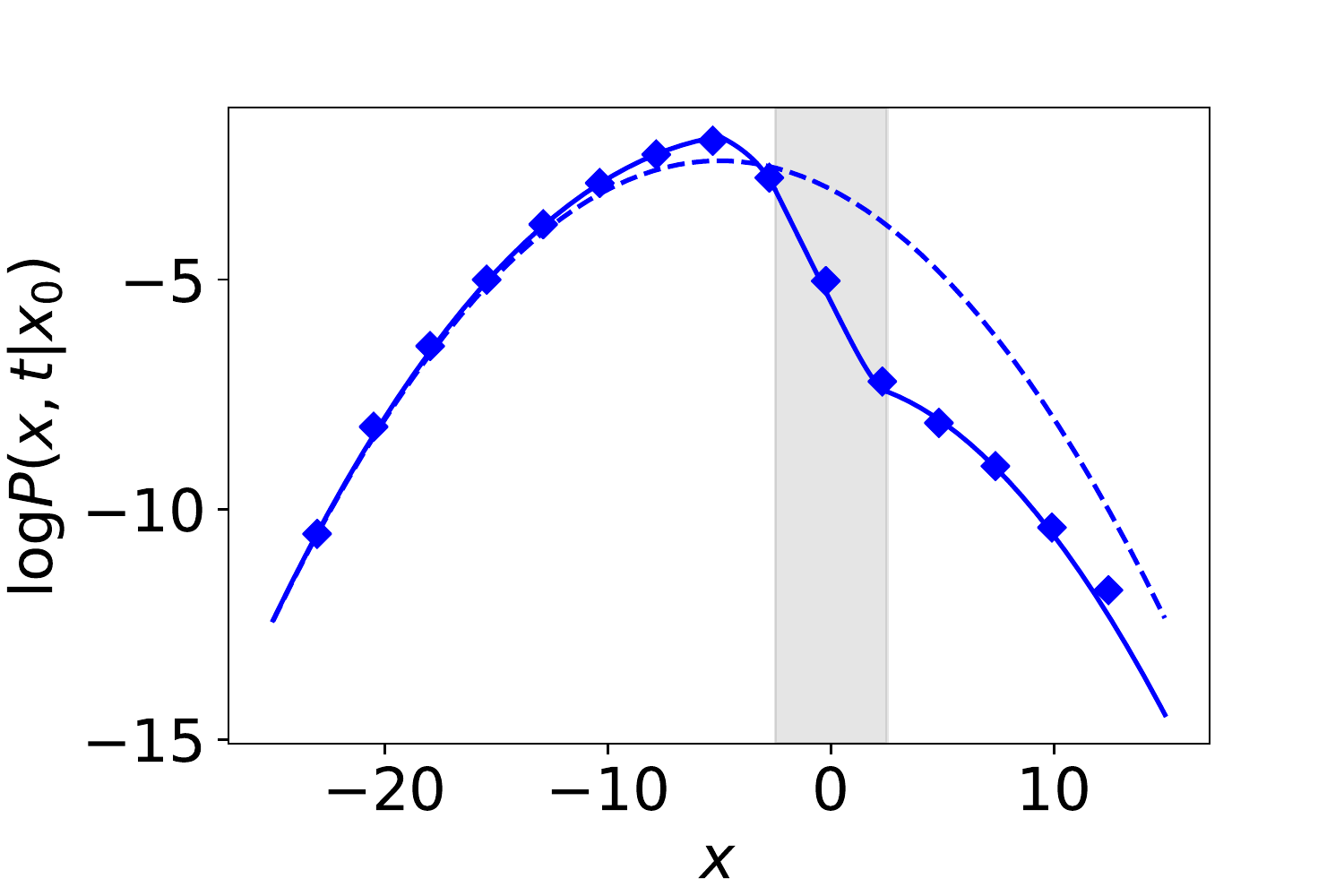}
\caption{\textbf{Brownian yet non-Gaussian diffusion induced by resetting.} Probability density $P(x,t|x_0)$ for a Brownian particle to be at position $x$ at time $t=10$ with $D=1$, $r=1.0$, $a=2.5$, $x_0=x_r=-5$, for the model in Fig.~\ref{fig:Deff1}a: numerical simulations (symbols), 
numerical inverse  Laplace transform of Eq.~\eqref{eq:PS1} (solid line), and exact (Gaussian) distribution in the absence of resetting, i.e. $r=0$ (dashed line).
 The grey shaded area highlights the interval where resetting occurs. The numerical simulations were done using the Euler numerical integration scheme with time step $\Delta t =0.05$, and  $N=10^5$ runs.}\label{fig:3}
\end{figure}

We now consider an
extension of the preceding problem, where a particle undergoing
evaporation is instantaneously reset to a given position $x_r$, see  Fig.~\ref{fig:Deff1}a for an illustration. The dynamics of the probability density $P_t(x)$ of
the particle position is described by Eqs.~\eqref{ME} and~\eqref{rx}
with $\sigma_c=1$, for which we derive an analytical solution in the
Laplace domain (see~Appendix~\ref{App_1}). We
point out two main features of the resulting distribution: 
the existence of a cusp at $x_r$~\cite{evans2011diffusion,Evans_2014} at
all times, and the absence of a stationary state due to the long-time prevalence of diffusion over resetting. Motivated by this observation, we investigate the long-time behaviour of the particle distribution moments. In particular, we quantify the drift and the
amplitude of fluctuations via the mean position $\langle x(t)\rangle$
and the variance $\sigma^2(t)\equiv \langle x^2(t)\rangle-\langle
x(t)\rangle^2$. At short times, until resetting kicks in, $\langle x(t)\rangle$ is constant as in Brownian diffusion.
At longer times, $\langle x(t) \rangle$ grows $\propto\sqrt{t}$ as
\begin{equation}
\langle x(t)\rangle=\sign(y_r)\sqrt{4D t/\pi}\phi(y_r,\rho)+O(t^0),
\label{eq:5}
\end{equation}
where
\begin{equation}
\begin{aligned}
\phi(y_r,\rho)&\equiv\begin{cases}
\tanh\left(\rho\right)\frac{\tanh\left(\rho\right)+\rho(|y_r|-1)}{\rho\tanh\left(\rho\right)(|y_r|-1)+1}&\!\!\!
\! \mbox{for}\, |y_r|>1,\\
\tanh\left(\rho\right)\tanh\left(|y_r|\rho\right)& \!
\!\!\!\mbox{for}\,|y_r|<1,
\end{cases}\\
\end{aligned}
\label{eq:xfull}
\end{equation}
with $y_r$ and $\rho$ given by Eq.~\eqref{eq:dimensionless}. 
Notably, a similar drift emerges also without resetting by
introducing a reflecting boundary (RB) at $x=0$. In this case, the
average particle position at long times is ${\langle x(t)\rangle_{\rm
RB}=\sign(x_0)\sqrt{4Dt/\pi}+O(t^0)}$, where the initial position $x_0$ of the particle plays formally the role of $x_r$ in Eq.~\eqref{eq:5}.
Because $|\phi(y_r,\rho)|<1$,  resetting acts as a
weaker reflecting boundary, whose efficiency depends on $\rho$ and $y_r$.
Due to the recurrence property of one-dimensional Brownian motion, even
for $|y_r|\gg 1$, a fraction of particles reach the resetting region and
is pushed back to $x_r$: the larger $x_r$ is, the longer the time
the particles spend in the half plane containing $x_r$ is. Accordingly, the
maximum value $\phi(y_r,\rho)=1$, corresponding to a reflecting boundary, is attained for  $|y_r|\rightarrow\infty$.

\begin{figure*}
\includegraphics[width=\textwidth]{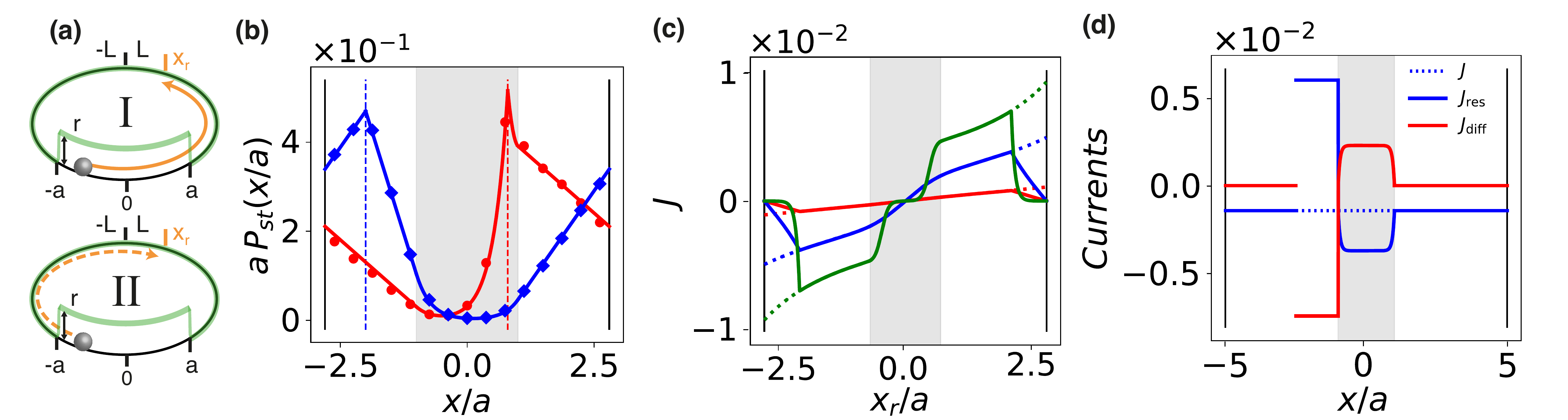}
\caption{\textbf{Diffusion with stochastic resetting on a ring}  according to different protocols (resetting ``gauge invariance''). \textbf{(a)} Sketch of two resetting protocols: the particles are reset to a fixed destination $x_r>0$ at a rate $r$ only when located in the window $[-a,a]$. Top (model I): resetting    moves the particle along the path that does not cross the endpoints $\pm L$. Bottom (model II): resetting occurs via the  path of minimal distance between the positions before and after resetting.  \textbf{(b)} Numerical results for the stationary spatial distribution  for $y_r=x_r/a=-2$ (blue diamonds) and $y_r=0.8$ (red circles), compared with the corresponding analytical predictions (solid lines, see Appendix~\ref{a:PBC}) with $D=50$, $r=1$,  $a=25$, $L=70$. \textbf{(c)} Stationary total particle current $J$ given by the right-hand side of Eq.~\eqref{eq:rescurr} for model I (dotted lines) and II (solid lines)  as a function of  the normalized resetting destination $y_r$, for  different resetting rates: $r=0.01$ (red), $0.1$ (blue) and  $10$ (green) with $D=5$, $a=7.5$, $L=30$. (d) Space-dependent decomposition of the total stationary current $J(x)= J_{\rm diff}(x)+J_{\rm res}(x)$ (dotted line) as a function of the normalized position $y=x/a$ for the model II: diffusive current $J_{\rm diff}(x)$ (blue line) and resetting current $J_{\rm res}(x)$ (red line). With $\rho=15.0$, $y_r=-2.5$ and $l=L/a=5$. A discontinuous jump of both $J_{\rm diff}(x)$ and $J_{\rm res}(x)$ occurs at $x=x_r$.  In all the panels, numerical results are obtained from $N=10^5$ simulations with time step $\Delta t =0.1$ while the gray area corresponds to the resetting region.}
\label{fig:PBC2}
\end{figure*}

The analytical expression for the variance turns out to have a long-time
behavior $\sigma^2(t)\propto t$. Accordingly, these particles diffuse without displaying a Gaussian  distribution, a
phenomenon that has recently attracted considerable attention in statistical physics known as {\em Brownian yet non-Gaussian}
diffusion~\cite{chubynsky2014diffusing,chechkin2017brownian,sposini2018random,miotto2019length}, see Fig. \ref{fig:3}.
 At long times, we
may define an effective diffusion coefficient $D_{\rm
eff}\equiv\lim_{t\rightarrow\infty} \sigma^2(t)/ 2t$, which generically differs from the diffusion coefficient
$D$~\cite{reimann2001giant,pietzonka2016universal}. For this model, we
have
\begin{equation}\label{eq:defffull}
\frac{D_{\rm eff}}{D}=1-\frac{2}{\pi}\phi^2(y_r,\rho),
\end{equation}
which depends only on the dimensionless variables $y_r$ and $\rho=a\sqrt{r/D}$ in Eq. \eqref{eq:dimensionless}, which encode the actual dependence on the width $a$ and the ratio $r/D$.
An interesting feature of $D_{\rm eff}$ is that it is bounded from above
by its maximum value $D_{\rm eff}^{\rm max}=D$, which is obtained under
symmetric resetting  $y_r=0$, while its minimum value  is attained for large $|y_r|$. Accordingly, we have
\begin{equation}\label{eq:Deff}
1-\frac{2}{\pi}< \frac{D_{\rm eff}}{D}\leq 1 .
\end{equation}
Resetting prevents the particles diffusing freely, implying $D_{\rm
eff}\leq D$. On the other hand, Eq.~\eqref{eq:Deff} reveals the existence of a lower bound $D_{\rm eff}^{\rm min}=D(1-2/\pi)$ for the effective diffusion coefficient, which  corresponds to the value for a Brownian particle diffusing near a reflecting wall. 
Figure~\ref{fig:Deff1}b confirms  the lower bound in Eq.~\eqref{eq:Deff} with numerical simulations of the present model, for various values of the model parameters. 
In addition, we have checked numerically that the same qualitative behaviour, i.e., a \textit{Brownian yet non-Gaussian} diffusion, occurs even when the resetting profile is made by the union of disjoint intervals of various lengths (data not shown). Moreover, one can recover the correct analytical predictions for the delta-like resetting rate $r_c(x)=r_0\delta(x)$ by setting $r=r_0/2a$ in Eqs.~\eqref{ME} and\eqref{rx} and by taking the limit $a\rightarrow0$.

\section{Resetting gauge invariance on a ring}\label{Sec:4}

Let us now consider the Brownian particle diffusing on the segment
$(-L, L)$ with periodic boundary conditions (e.g., on a ring of perimeter
$2L$) which, with a constant rate $r$, experiences resetting to a point
$x_r$ when it is within the interval $(-a,a)$ with $a<L$, see Fig.
\ref{fig:PBC2}a for an illustration. Periodic boundary conditions of this model ensure the
existence of a stationary probability distribution for the particle
position with a cusp, (global maximum) at $x_r$ \cite{evans2011diffusion,Evans_2014}, as shown in
Fig.~\ref{fig:PBC2}b, where we compare analytical (see Appendix \ref{a:PBC}) and numerical results for $P_t(x)$ for various values of the relevant parameters. 
Importantly, in order to characterize resetting on the ring, one needs to specify the physical direction of the particle flux arising from it. In this framework, it is often assumed that particles, once reset, reach $x_r$ instantaneously, i.e., in a teleporting fashion, without  specifying \textit{how} this actually occurs. For example, on a ring, particles may reset by moving always clockwise, always counterclockwise or in both directions.  The one-time statistics of the particle position is a physical observable quantity whose distribution $P(x,t|x_0)$  is actually independent of how resetting occurs, as long as it is instantaneous. On the other hand, local and conserved particle currents arise in ring-like geometries and their values depend on the specific resetting rule. 
In this spirit, it is natural to define a {\em resetting current}
$J_{\rm res}(x,t)$ by integrating the right-hand side of Eq.~\eqref{ME}:
\begin{equation}
\begin{aligned}
J_{\rm res}(x,t)-J_{\rm res}(-L,t)&=\int_{-L}^x\mathrm{d}y\,r_c(y)P(y,t|x_0)\\
&-\theta(x-x_r)R_t,\label{eq:rescurr}
\end{aligned}
\end{equation}
where the spatial constant $J_{\rm res}(-L,t)$, up to which the
space-dependence of $J_{\rm res}(x,t)$ is defined, stems from the gauge freedom in the choice of the specific protocol according to which resetting actually occurs.  
Thus, the total probability current  $J(x,t)\equiv J_{\rm
res}(x,t)+J_{\rm diff}(x,t)$ with $J_{\rm diff}(x,t)=-D\partial_x
P(x,t|x_0)$, obeys the continuity (Fokker-Planck) equation~\eqref{ME}, which can be written as 
$\partial_t P(x,t|x_0)=-\partial_x J(x,t)$. Note that as long as
resetting is instantaneous, the particle diffusive dynamics is insensitive to how
resetting occurs. Consistently, $P(x,t|x_0)$ $-$and hence $J_{\rm
diff}(x,t)$$-$ does not depend on the details of the resetting
protocol, while $J_{\rm res}$ $-$and hence $J$$-$ does depend by an overall
possibly time-dependent spatial constant. In the stationary state, the
total current $J(x,t)$ either vanishes, as it happens in infinite
systems, or becomes space-independent, while $J_{\rm res}$ and $J_{\rm
diff}$ are generically space-dependent, as shown below (see
Appendix~\ref{a:PBC}). Thus, we have gauge invariance in both stationary and
non-stationary regimes: many different currents, corresponding to the different reset protocols, refer to the same spatial probability distribution.

We illustrate the resetting gauge invariance for two systems with the same geometry but
with different resetting protocol: (model I, Fig.~\ref{fig:PBC2}a, top)  resetting along the path
that never crosses the endpoints $\pm L$, i.e., $J_{\rm res}(-L)=0$, and
(model II, Fig.~\ref{fig:PBC2}a, bottom): resetting according to the minimal path protocol, i.e.,  the particle, resetting at $x_{\rm res}$, travels the path of minimum distance $\Delta
x_{\rm min}=\min(|x_r-x_{\rm res}|,2L-|x_r-x_{\rm res}|)$. For these two resetting mechanisms, we report in Fig.~\ref{fig:PBC2}b the stationary density
$P_{\rm st}$  as a function of the position $x$ along the
ring, and in in Fig.~\ref{fig:PBC2}c the total stationary current $J$  as a function of the coordinate
$x_r$ of the resetting point.  While the
stationary spatial distribution is insensitive to the resetting protocol
(see Fig.~\ref{fig:PBC2}b), the stationary total current depends strongly on
the resetting protocol (see Fig.~\ref{fig:PBC2}c).    The latter displays robust qualitative features in its dependence on the resetting destination $x_r$, see Fig.~\ref{fig:PBC2}c: with $x_r$ outside (inside)  the resetting region compared to the typical values outside it, 
i.e., $|x_r|>a$ ($|x_r|<a$), $J$ increases (decreases) upon increasing
(decreasing) the resetting rate $r$, because resetting induces particles
to be concentrated in a region with low (high) local resetting rate. As
a result, the current is exponentially suppressed upon increasing $r$
inside the resetting region, inducing a discontinuity at $x=\pm a$  for $r\rightarrow\infty$. Notably, our finding rationalizes the emergence of a cusp in the stationary distribution at $x_r$, as the resetting current is discontinuous also in $x_r$ due to the imbalance of resetting fluxes (Fig.~\ref{fig:PBC2}d).

\begin{figure}
\includegraphics[scale=0.38]{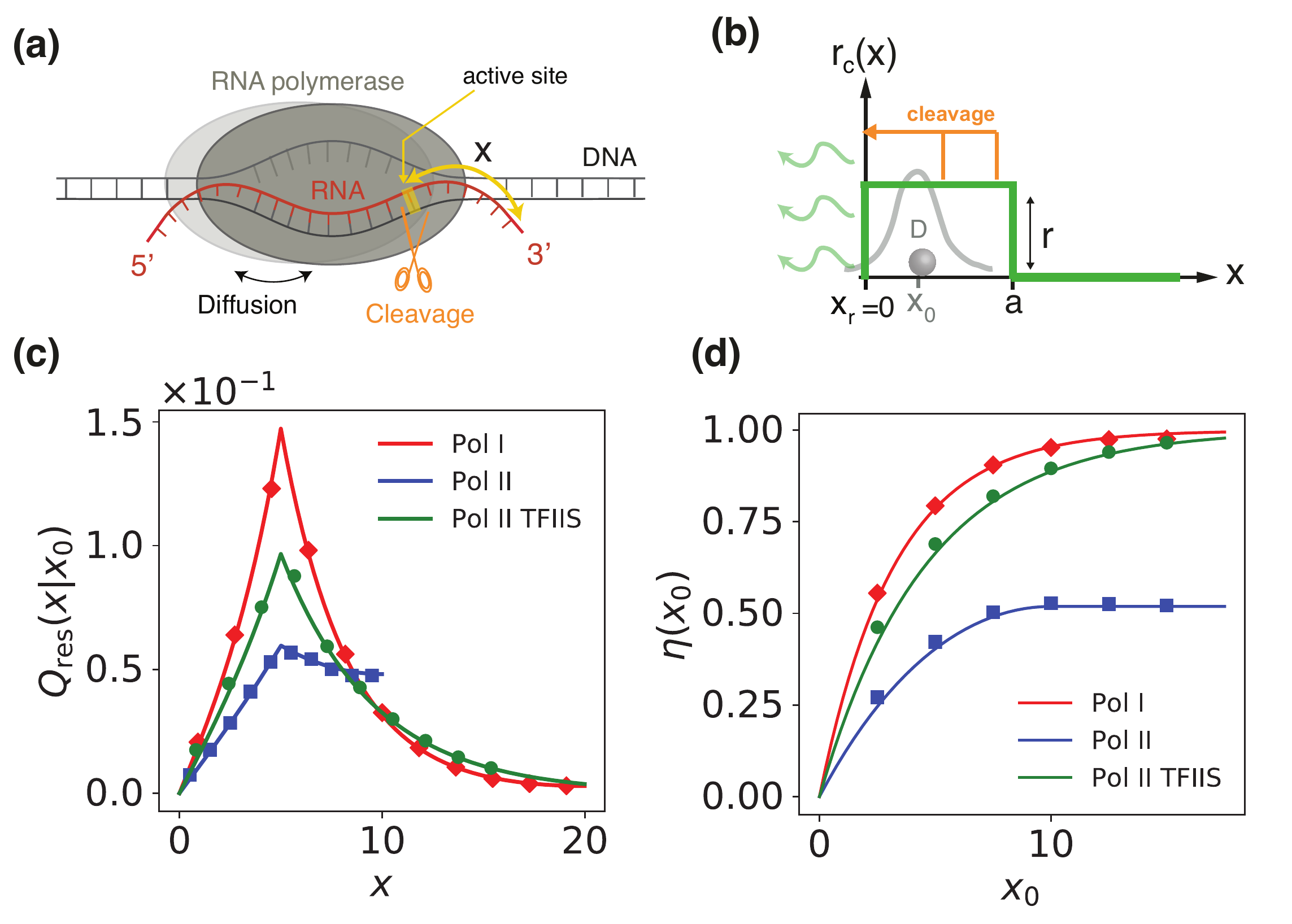}
\caption{\textbf{Modelling backtrack recovery of RNA polymerase}.  \textbf{(a)} Illustration of an RNA polymerase (grey circle) diffusing along a DNA template (black ladder). During a transcriptional pause, the polymerase cannot resume RNA polymerization because its active site (yellow box) $-$where new nucleotides are added$-$ is blocked by backtracked RNA (red line). Polymerases "recover" from the backtracking state when the distance  $x\geq 0$ ("backtrack depth", yellow arrows) between the active site and the 3'-end of the RNA vanishes. Recovery from backtracking is due to Brownian diffusion of the polymerase with diffusion coefficient $D$  (grey arrows) or to the action of active processes that allow cleavage (orange scissors) of backtracked RNA  of length up to a finite length $a\geq 0$ and at random times with constant rate $r$~\cite{lisica2016mechanisms}.   \textbf{(b)} Model for the evolution of the backtrack depth $x$ from an initial value $x_0$.  Polymerases recover by reaching the absorbing state $x=0$ either by diffusion or by resetting, representing RNA cleavage~\cite{lisica2016mechanisms,roldan2016stochastic}. \textbf{(c)} Distribution of the length of cleaved RNA for Pol I,  Pol II, and Pol II TFIIS with initial backtrack depth $x_0=5$ (nucleotides). \textbf{(d)} Cleavage (resetting) efficiency as a function of the initial backtrack depth. The numerical prediction for $\eta_{\rm res}(y_0)\equiv\eta(x_0=a\,y_0)$ has been already reported in Ref. \cite{lisica2016mechanisms}. In both (b) and (c) panels, the symbols are obtained from numerical simulations and the solid lines are theoretical predictions given by Eqs.~\eqref{eq:distlength2} in (c) and~\eqref{eq:reseff2} in~(d). Values of the parameters $D$, $r$ and $a$ can be found in Appendix~\ref{App:Pol}. Simulations were done with time step $\Delta t =0.1$ for panel (c) and $\Delta t =0.025$ for panel (d) with number of simulations $N=10^5$.} \label{fig:Pol} 
\end{figure}
  
\section{Application to RNA polymerase}\label{Sec:5}
We now apply our theory  (see Appendix \ref{App:Pol}) to a biophysical model 
describing the fluctuating motion of RNA polymerases along a DNA template during transcriptional pauses, introduced in Ref.~\cite{lisica2016mechanisms}. 
Figure~\ref{fig:Pol}a sketches two recovery mechanisms that can be employed by an RNA polymerase enzyme to recover from the inactive state ("backtracking"): (i) Brownian diffusion due to thermal  fluctuations,  and (ii) active cleavage of the backtracked RNA induced by chemical reactions. This dynamics can be modeled as shown in Fig.~\ref{fig:Pol}b, as confirmed by {\em in vitro} single-molecule experimental data obtained for yeast {\em S. Cerevisiae}~\cite{lisica2016mechanisms}. The model describes the evolution of  the {\em backtrack depth} $x\geq 0$, i.e., the spatial distance  between the active site of the polymerase and the 3' end of the backtracked RNA. The dynamics consists of diffusion in $d=1$ starting from the initial value $x_0>0$ with an absorbing boundary in the origin corresponding to the return to the RNA polymerization state. Cleavage of backtracked RNA is modelled by a sudden
jump (i.e., resetting) of the backtrack depth  $x\to x_r=0$  at a rate $r$ from the region $(0,a)$; note that resetting is equivalent here to evaporation. The choice of a finite interval for resetting is motivated by previous experimental observations which suggested that  polymerases can only cleave RNA transcripts of a finite length~\cite{kuhn2007functional,engel2013rna},  and has been  tested in recent single-molecule experiments~\cite{lisica2016mechanisms}.

The probability density $P_{a}(x,t|x_0)$ for the backtrack depth $x$, reported in Appendix~\ref{App:Pol}, is computed by applying the so-called image method~\cite{redner_2001} to the probability density $P(x,t|x_0)$ for the Brownian particle resetting in the inteval $[-a,a].$ To account for the absorbing boundary at $x_a=0$, we place a  "negatively charged" source of particles in $-x_0$ resetting at rate $r$ in the complementary interval $(-a,0)$ to the same $x_r=0$, leading to $P_{a}(x,t|x_0)=P(x,t|x_0)-P(x,t|-x_0)$. Accordingly, the presence of the absorbing boundary prevents the system from reaching a non-trivial stationary state and therefore the existence of non-vanishing currents at long times.

We study  two statistical properties of the RNA polymerase backtracking 
for which no analytical results have been reported so far. Firstly, we focus on the distribution of the length of cleaved RNAs. In the model, this quantity corresponds to the probability density $Q_{\rm res}(x|x_0)$  of particles, with initial point $x_0$, that are absorbed at any time from the position $x$ through resetting only.
In terms of the dimensionless quantities in Eq. \eqref{eq:dimensionless}, it is given by
\begin{equation}\label{eq:distlength2}
\begin{aligned}
Q_{\rm res}&(x=a\,y|x_0=a\,y_0)=\frac{\rho\sinh(y_<\rho)}{a\cosh\rho}\\
&\times\left\{1-\theta(1-y_0)\left[1-\cosh\left((1-y_>)\rho\right)\right]\right\},
\end{aligned}
\end{equation}
with  $y_>\equiv\max(y,y_0)$, $y_<\equiv\min(y,y_0)$; note that in the region $x>a$ ($y>1$) $Q_{\rm res}$ vanishes due to the absence of resetting.
Secondly, we quantify the overall efficiency of cleavage in recovery by defining  the ``resetting efficiency"  $\eta_{\rm res}(y_0)\leq 1$ as the fraction of polymerases that recover from any $x\leq a$ and at any time $t>0$ due to RNA cleavage (resetting).  Its analytical expression can be found by using  Eq.~\eqref{eq:distlength2} (see Appendix~\ref{App:Pol}):
\begin{eqnarray}
&&\eta_{\rm res}(y_0)=\label{eq:reseff2}\\
&&1-\frac{\cosh\left((1-y_0)\rho\right)+\theta(y_0-1)\left[1+\cosh\left((1-y_0)\rho\right)\right]}{\cosh\rho}.\nonumber
\end{eqnarray}
Notably, the distribution~\eqref{eq:distlength2} of cleaved RNA length and the efficiency~\eqref{eq:reseff2} of RNA cleavage  obey  universal scaling laws in terms of the parameter  $\rho$ and $y_0$ in Eq.~\eqref{eq:dimensionless}. We demonstrate these results in Fig.~\ref{fig:Pol} with numerical simulations  using values of the parameters which were measured for the enzymes Pol I, Pol II and Pol II omplementha TFIIS from yeast {\em S. Cerevisiae}~\cite{lisica2016mechanisms}. Interestingly, the distributions of the length of the cleaved RNA in Fig.~\ref{fig:Pol}c display a cusp at the initial position and exponential tails. $Q_{\rm res}(x|x_0)$ is positive on a broader range of positions for Pol I and Pol II-TFIIS,
because these enzymes can cleave RNA of lengths $a$ larger than Pol II. 
On the other hand, the cleavage efficiency, ${\eta_{\rm res}(y_0)\equiv \eta(x_0=a\,y_0)}$ in Fig.~\ref{fig:Pol}d, increases monotonously with the initial backtrack depth $x_0$ for all the enzymes. Notably, Pol II maximum cleavage efficiency attained for deep backtracks is $\sim$50\% for Pol II and twice larger and almost $\sim$100\% for both Pol I and Pol II-TFIIS. The similarities in the cleavage efficiency  of Pol I and Pol II-TFIIS may stimulate further research on evolutionary-conserved  performance between different types of transcription enzymes.

\section{Discussion}\label{Sec:6}
Our work provides analytical and
numerical insights into the particle currents emerging in the presence of control mechanisms
(evaporation and resetting) on an otherwise unbiased Brownian diffusion, for various boundary conditions. For open boundary conditions, we have
shown that resetting the particle position at stochastic times to a
prescribed location leads to Brownian yet non-Gaussian diffusion. On the other hand, periodic boundary conditions induce
stationary particle current that can be decomposed as the sum of
diffusive and resetting fluxes. Such a current could be used, e.g., to exert a force on an external load, as in the case of Brownian motors. For
ring-like geometries, we have proved a resetting gauge invariance of the
distribution with respect to the resetting protocol (direction), resulting into different
particle currents. Finally, we have applied our findings to  a biophysical model, deriving  analytical predictions that involve the efficiency of  RNA polymerase backtracking. Our work indicates new avenues for understanding the nonequilibrium features of resetting, e.g., in the study of optimization of resetting pathways for efficient particle transport. Furthermore, we expect that  our formalism could be extended to shed light on various biophysical problems described by one-dimensional diffusions with suitable boundary and/or resetting conditions, such as microtubule dynamics~\cite{nayak2020comparison,prelogovic2019pivot}, molecular motors~\cite{julicher1997modeling,keller2000mechanochemistry,guillet2019extreme}, single-file diffusion of water in carbon nanorings~\cite{mukherjee2010single},  or polymer translocation through nanopores~\cite{abdolvahab2011first,abdolvahab2011sequence}.


\begin{acknowledgments}
S.G. acknowledges support from the Science
and Engineering Research Board (SERB), India under SERB-TARE scheme Grant No. TAR/2018/000023 and SERB-MATRICS scheme Grant No. MTR/2019/000560. He also thanks ICTP -- The Abdus Salam International Centre for Theoretical Physics, Trieste, Italy for support under its Regular Associateship scheme.
\end{acknowledgments}

\newpage
\onecolumngrid
\section*{APPENDIX}
\appendix

\section{GENERAL EXPRESSIONS FOR $\tilde{P}_{\rm nc}$ AND $\tilde{P}$}\label{Appendix:P}

In this Appendix we report the solution for the Laplace transform of Eq. \eqref{ME} for $P(x,t|x_0,x_r)$, the probability distribution of the Brownian particle position in one dimension, initially at $x_0$, resetting to $x_r$ according the space-dependent resetting rate $r_c(x)$. Although the framework that we will present applies to any space-dependent resetting rate, we will focus on the case $r_c(x)=r\,\theta(a-|x|)$, i.e., non vanishing within a segment of width $2a$ and constant rate $r$.
More generally, we exert an external control on the system, parametrized by $\sigma_c$, corresponding to the identification of $r_c(x)$ as a resetting rate ($\sigma_c=1$) or as an evaporation rate ($\sigma_c=0$). In this picture, we denote by $P_{\rm nc}(x,t|x_0)$ the probability density of no-control, i.e., the probability that no reset/evaporation has occurred up to time $t$ in the space interval $[x,x+\mathrm{d}x]$ is given by $P_{\rm nc}(x,t|x_0)\mathrm{d}x$.
In general, $P(x,t|x_0,x_r)$ depends both on the initial position $x_0$ and the resetting point $x_r$; in what follows, we denote this quantity by $P(x,t|x_0)$ whenever $x_0\equiv x_r.$
Note that in the case of evaporation, $\sigma_c=0$,   the probability distribution $P(x,t|x_0,x_r)$ coincides with $P_{\rm nc}(x,t|x_0)$, because the only existing particles are those that have not experienced evaporation.

Analogously, following Ref.~\cite{path_integral}, $P_{\rm nc}(x,t|x_0)$ corresponds, in the case, $\sigma_c=1$, of resetting to the probability density $P_{\rm no\,res}(x,t|x_0)$ for particles with initial and final position $x_0$ and $x$, respectively, not to reset in the time interval $(0,t).$ We now assume $P_{\rm nc}(x,t|x_0)$ given and we fix $\sigma_c=1$, focusing on the model with resetting.
In terms of $P_{\rm nc}(x,t|x_0)$ the probability $P_{\rm res}(t|x_0)$ of first reset at time $t$ can be evaluated as follows
\begin{equation}\label{eq:pres}
P_{\rm res}(t|x_0)=\int_{-\infty}^{+\infty}\mathrm{d}y\,r_c(y)\,P_{\rm nc}(y,t|x_0,0).
\end{equation}
Once $P_{\rm nc}(x,t|x_0)$ and $P_{\rm res}(t|x_0)$ are known we can construct $P(x,t|x_0,x_r)$ by means of renewal theory \cite{path_integral}:

\begin{equation}\label{eq:general_sol}
\begin{aligned}
P(x,t|x_0,x_r)&=P_{\rm nc}(x,t|x_0)+\int_0^t\mathrm{d}\tau\int_{-\infty}^{+\infty}\mathrm{d}y\,r_c(y)P(y,t-\tau|x_0)P_{\rm nc}(x,t|x_r,t-\tau)\\
&=P_{\rm nc}(x,t|x_0)+\int_0^t\mathrm{d}\tau R(t-\tau|x_0)P_{\rm nc}(x,\tau|x_r),
\end{aligned}
\end{equation}
where the flux $R(t|x_0)$ of particles resetting at time $t$ is defined as
\begin{equation}
R(t|x_0)\equiv\int_{-\infty}^{+\infty}\mathrm{d}y\,r_c(y)P(y,t|x_0),
\end{equation}
henceforth indicated by $R_t$.
The first term on the right-hand side of Eq. \eqref{eq:general_sol} corresponds to particles that never reset while the second corresponds to trajectories where particles reset for the last time in any position $y$ at any intermediate time $t-\tau$ and then restart from $x_r$, subsequently diffusing to $x$ without undergoing resetting within the time interval $(t-\tau,t)$.
Operatively, the appearance of $P(x,t|x_0,x_r)$ on both sides of Eq. \eqref{eq:general_sol} makes its solution natural in terms of the Laplace transform; in what follows we use a tilde to denote the Laplace transform of a function $\tilde{f}(s)\equiv\int_0^\infty\mathrm{d}t\,e^{-st}f(t)$.
In particular, Eq. \eqref{eq:general_sol} reduces to

\begin{equation}\label{eq:LTP}
\begin{aligned}
\tilde{P}(x,s|x_0,x_r)=\tilde{P}_{\rm nc}(x,s|x_0)+\tilde{R}(s|x_0)\tilde{P}_{\rm nc}(x,s|x_r),
\end{aligned}
\end{equation}
where $\tilde{P}_{\rm nc}$ appears both with initial point $x_0$ and $x_r$. 
By multiplying Eq. \eqref{eq:general_sol} by $r_c(x)$ and integrating over $x$ one gets

\begin{equation*}
R(t|x_0)=P_{\rm res}(t|x_0)+\int_0^t\mathrm{d}\tau\,R(t-\tau|x_0)P_{\rm res}(\tau|x_r),
\end{equation*}
whose Laplace transform is given by

\begin{equation}\label{eq:LTR}
\tilde{R}(s|x_0)=\tilde{P}_{\rm res}(s|x_0)+\tilde{R}(s|x_0)\tilde{P}_{\rm res}(s|x_r).
\end{equation}
By combining Eqs. \eqref{eq:LTP} and \eqref{eq:LTR} we derive a closed expression for $\tilde{P}(x,s|x_0,x_r)$:

\begin{equation}\label{psxr}
\begin{aligned}
\tilde{P}&(x,s|x_0,x_r)=\tilde{P}_{\rm nc}(x,s|x_0)
+\frac{\tilde{P}_{\rm res}(s|x_0)}{1-\tilde{P}_{\rm res}(s|x_r)}\tilde{P}_{\rm nc}(x,s|x_r);
\end{aligned}
\end{equation}
if $x_0= x_r$ Eq. \eqref{psxr} reduces to $\tilde{P}(x,s|x_0)=\frac{\tilde{P}_{\rm nc}(x,s|x_0)}{1-\tilde{P}_{\rm res}(s|x_0)}$, allowing to recast the same Eq. \eqref{psxr} as

\begin{equation}\label{eq:Pgen2}
\begin{aligned}
\tilde{P}&(x,s|x_0,x_r)=\left[1-\tilde{P}_{\rm res}(s|x_0)\right]\tilde{P}(x,s|x_0)+\tilde{P}_{\rm res}(s|x_0)\tilde{P}(x,s|x_r).
\end{aligned}
\end{equation}
We now outline the procedure  used throughout the paper to compute $\tilde{P}(x,s|x_0,x_r)$:

\begin{itemize}
\item[1)] First,  compute $\tilde{P}_{\rm nc}(x,s|x_0)$ from the Laplace transform of Eq.\eqref{ME} with $\sigma_c=0$, i.e.,
\begin{equation}\label{LaplaceSE}
\begin{aligned}
&D\frac{\partial^2 \tilde{P}_{\rm nc}(x,s|x_0)}{\partial x^2}-\left(s+r_c(x)\right)\tilde{P}_{\rm nc}(x,s|x_0)=-\delta(x-x_0)
\end{aligned}
\end{equation}
with the proper boundary conditions.
\item[2)] Evaluate $\tilde{P}_{\rm res}(s|x_0)$ using Eq. \eqref{eq:pres}.
\item[3)] Use Eq. \eqref{psxr} in order to compute $\tilde{P}(x,s|x_0,x_r).$
\end{itemize}

As anticipated, we will compute explicitly $\tilde{P}(x,s|x_0,x_r)$ for $r_c(x)=r\,\theta(a-|x|)$, which requires the separate analysis of the two cases $|x_0|<a$ and $|x_0|>a.$

\section{BROWNIAN TUNNELLING WITH RESETTING}\label{App_1}

\subsection{Case with $x_0<-a$}\label{x0<-a}

Due to the presence of a piece-wise resetting rate in Eq. (\ref{LaplaceSE}), we have to consider separate contributions  to $\tilde{P}_{\rm nc}(x,s|x_0)$ for any of the regions delimited by the boundary points $\{\pm a, x_0\}$:
\begin{equation}\label{eq:gensol}
\begin{aligned}
&\tilde{P}_{\rm nc}(x,s|x_0)=\begin{cases}
A(s,x_0)\,e^{x\mu} & \mbox{for}\,\, x<x_0,\\
B_1(s,x_0)\,e^{x\mu}+B_2(s,x_0)\,e^{-x\mu} & \mbox{for}\,\, x\in(x_0,-a),\\
C_1(s,x_0)\,e^{x\nu}+C_2(s,x_0)\,e^{-x\nu}& \mbox{for}\,\,x\in(-a,a),\\
E(s,x_0)\,e^{-x\mu} & \mbox{for}\,\, x>a,
\end{cases}
\end{aligned}
\end{equation}
where we define 

\begin{equation}\label{eq:munu}
\begin{aligned}
\mu(s)\equiv\sqrt{\frac{s}{D}}\,\,\,\,\,\,\,\,\,\nu(s)&\equiv\sqrt{\frac{r+s}{D}},
\end{aligned}
\end{equation}
and the coefficients $A$, $B_{1,2}$, $C_{1,2}$, and $E$ are fixed as specified below. 
The non-divergence of the solution is ensured by requiring that the real part of $\mu$ is positive, $\text{Re}(\mu)>0$, this is equivalent to require that $\tilde{P}_{\rm nc}$ is defined on the entire complex $s$-plane except the negative real axis. We fix all the coefficients in Eq. \eqref{eq:gensol} by requiring the continuity of $\tilde{P}_{\rm nc}$ at the boundary points and the generic continuity of the first derivative in all points, except for $x_0$. In fact, given the presence of a Dirac delta in Eq. \eqref{LaplaceSE}, $\tilde{P}_{\rm nc}$ will present a discontinuity in the derivative according to

\begin{equation}\label{eq:derjump}
\partial_x\tilde{P}_{\rm nc}(x_0^+,s|x_0)-\partial_x\tilde{P}_{\rm nc}(x_0^-,s|x_0)=-\frac{1}{D}.
\end{equation}

This derives from integrating Eq. \eqref{LaplaceSE} around a small neighborhood of $x_0$ of radius $\epsilon>0$ and then taking the limit $\epsilon\rightarrow 0.$ Accordingly, the three conditions for the continuity of $\tilde{P}_{\rm nc}$ at the boundary points, the two conditions for the continuity of the first space derivative at $\pm a$ and Eq. \eqref{eq:derjump} fix the constants in the solution \eqref{eq:general_sol}.
The final expressions for $\tilde{P}_{\rm nc}$ read

\begin{small}
\begin{equation}\label{eq:Pnc1}
\begin{aligned}
&\tilde{P}_{\rm nc}(x,s|x_0)=\frac{1}{D}\left[(\mu^2+\nu^2)\sinh\left(2a\nu\right)+2\mu\nu\cosh\left(2a\nu\right)\right]^{-1}\\
&\\[-8pt]
&\begin{cases}
\left\{e^{\mu(x-x_0)}\left[\mu^2 \sinh\left(2a\nu\right)+\nu\mu\cosh\left(2a\nu\right)\right]-\frac{r}{D}e^{\mu(x+a)}\sinh\left(2a\nu\right)\sinh\left(\mu(a+x_0)\right)\right\}/\mu & \mbox{for}\,\, x\le x_0,\\
&\\[-8pt]
\left\{e^{-\mu(x-x_0)}\left[\mu^2\sinh\left(2a\nu\right)+\nu\mu\cosh\left(2a\nu\right)\right]-\frac{r}{D}e^{\mu(x_0+a)}\sinh\left(2a\nu\right)\sinh\left(\mu(a+x)\right)\right\}/\mu & \mbox{for}\,\, x\in(x_0,-a),\\
&\\[-8pt]
e^{\mu(x_0+a)}\left[\nu\cosh\left(\nu(x-a)\right)-\mu\sinh\left(\nu(x-a)\right)\right]& \mbox{for}\,\, x\in[-a,a],\\
&\\[-8pt]
e^{\mu(2a+x_0-x)}\nu & \mbox{for}\,\, x>a;
\end{cases}
\end{aligned}
\end{equation}
\end{small}

\begin{figure*}
\includegraphics[width=\textwidth]{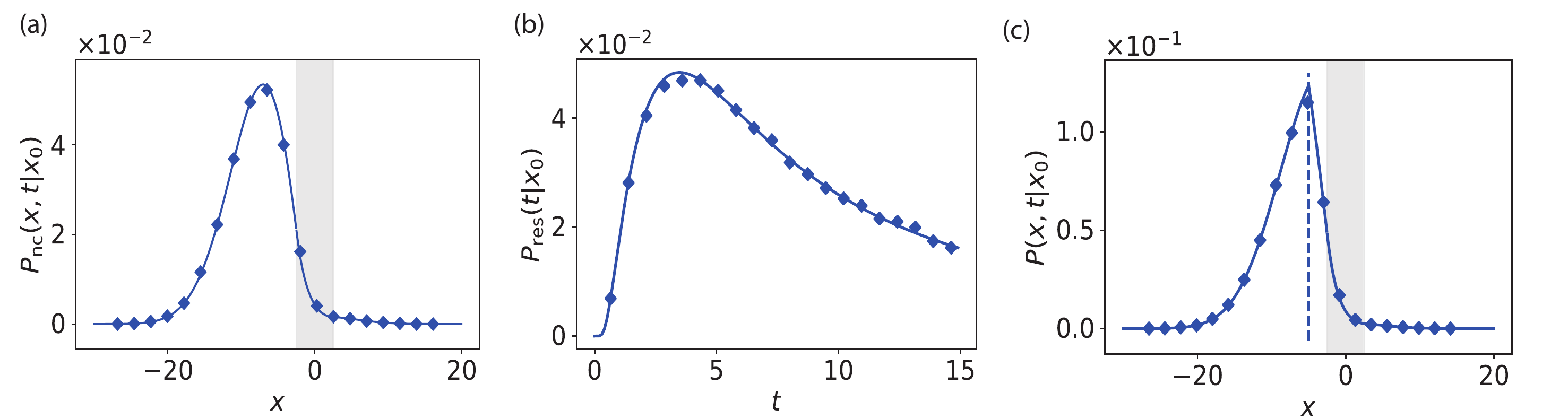}
\caption{\textbf{Statistics of the Brownian tunnelling:} (a)  Probability density $P_{\rm nc}(x,t|x_0)$ at time $t=15$ with $D=1$, $r=0.5$, $a=2.5$, $x_0=-5$. The solid line represents the inverse numerical Laplace transform of Eq. \eqref{eq:Pnc1} while the symbols correspond to the result of numerical simulations. The grey shaded area highlights the interval within which resetting occurs. (b) Probability of first reset time $P_{\rm res}(t|x_0)$ up to time $t=15$ with the same parameters as panel (a): comparison between inverse Laplace transform of Eq. \eqref{eq:Pres1} and numerical simulations. (c) Probability density $P(x,t|x_0)$ with $x_r=x_0$, indicated by the vertical dashed line. The solid line represents the inverse Laplace transform of Eq. \eqref{eq:PS1} while symbols indicate the results of numerical simulations. All numerical simulations were done using the Euler numerical integration scheme with time step $\Delta t =0.05$, and they were repeated $N=10^5$ times.} \label{fig:Pnc1}
\end{figure*}
Fig. \ref{fig:Pnc1}a shows a representative instance of $P_{\rm nc}(x,t|x_0)$. This probability distribution has no stationary distribution and "evaporation" prevents the conservation of probability.
In the case of resetting, one may directly compute the Laplace transform of the probability distribution for the first reset time through Eqs. \eqref{eq:pres} and \eqref{eq:Pnc1}, finding

\begin{equation}\label{eq:Pres1}
\tilde{P}_{\rm res}(s|x_0)=\frac{r\,e^{\mu(a+x_0)}}{D\nu}\frac{\sinh(a\nu)}{\nu\sinh(a\nu)+\mu\cosh(a\nu)},
\end{equation}
the inverse transform of which is represented in Fig. \ref{fig:Pnc1}b.
The long-time behavior of $P_{\rm res}$, see Ref. \cite{redner_2001}, can be extracted from the behavior for small $s$ of its Laplace transform,

\begin{equation}\label{eq:presexp}
\tilde{P}_{\rm res}(s|x_0)= 1+\sqrt{s}\left(\frac{a+x_0}{\sqrt{D}}-\frac{\coth(a\sqrt{\frac{r}{D}})}{\sqrt{r}}\right)+O(s),
\end{equation}
which, plugged in Eq. \eqref{psxr}, implies the absence of a stationary distribution for the process, being  $P_{\rm nc}(x,s|x_0)=O(1)$ as $s\rightarrow 0$, ${P_{\rm st}(x|x_0)=\lim_{s\rightarrow 0}s\frac{P_{\rm nc}(x,s|x_0)}{1-P_{\rm res}(s|x_0)}=0}$.\\
Moreover, apart for the zeroth-order term in Eq. \eqref{eq:presexp} which corresponds to the normalization of the probability, the small $s$ expansion shows the presence of half-integer powers which reflects the fact that the integer moments of $\tilde{P}_{\rm res}(s|x_0)$ are infinite: e.g., the average time that a particle takes to reset is infinite because of the existence of infinite diffusive paths extending towards increasingly negative positions. The long-time behavior $P_{\rm res}\sim t^{-\frac{3}{2}}$ for $t\rightarrow \infty$ also follows from this expansion.

Finally, according to Eq. \eqref{psxr}, using Eqs. \eqref{eq:Pnc1} and \eqref{eq:Pres1}, the Laplace transform of the probability distribution $P(x,t|x_0,x_r)$ for $x_r=x_0$ reads
\begin{small}
\begin{equation}\label{eq:PS1}
\begin{aligned}
&\tilde{P}(x,s|x_0)=\frac{\nu}{2D}\left\{\left[\mu\sinh\left(a\nu\right)+\nu\cosh\left(a\nu\right)\right]\left[\left(\nu^2-\frac{r}{D} e^{\mu(x_0+a)}\right)\sinh\left(a\nu\right)+\mu\nu\cosh\left(a\nu\right)\right]\right\}^{-1}\\
&\\[-8pt]
&\begin{cases}
\left\{e^{\mu(x-x_0)}\left[\mu^2 \sinh\left(2a\nu\right)+\nu\mu\cosh\left(2a\nu\right)\right]-\frac{r}{D}e^{\mu(x+a)}\sinh\left(2a\nu\right)\sinh\left(\mu(a+x_0)\right)\right\}/\mu & \mbox{for}\,\, x\le x_0,\\
&\\[-8pt]
\left\{e^{-\mu(x-x_0)}\left[\mu^2\sinh\left(2a\nu\right)+\nu\mu\cosh\left(2a\nu\right)\right]-\frac{r}{D}e^{\mu(x_0+a)}\sinh\left(2a\nu\right)\sinh\left(\mu(a+x)\right)\right\}/\mu & \mbox{for}\,\, x\in(x_0,-a),\\
&\\[-8pt]
e^{\mu(x_0+a)}\left[\nu\cosh\left(\nu(x-a)\right)-\mu\sinh\left(\nu(x-a)\right)\right]& \mbox{for}\,\, x\in[-a,a],\\
&\\[-8pt]
e^{\mu(2a+x_0-x)}\nu & \mbox{for}\,\, x>a.
\end{cases}
\end{aligned}
\end{equation}
\end{small}
Note that, as expected, for $r=0$ renders the Laplace transform of a Gaussian distribution with average $x_0$ and variance $2Dt$, i.e., $\tilde{P}(x,s|x_0)=e^{(-|x-x_0|\sqrt{s/D})}/(2\sqrt{s\,D}).$
The cusp for $x_r=x_0$ is the distinctive feature of resetting being present at all times. Moreover, for short and long times, it can be easily checked that the tails of the distribution at large values of $x$ are Gaussian, signaling that diffusion is the main mechanism driving the system. These properties are clearly displayed in Fig. \ref{fig:Pnc1}c which reports a snapshot of the probability density $P(x,t|x_0)$ showing the cusp in correspondence of $x_r$ (vertical dashed line) and the exponential suppression of the probability upon advancing within the resetting area (grey region) due to particles that reset. As a final remark, we point out that the solution for $x_0>a$ can be retrieved by that corresponding to $x_0<-a$ upon exchanging $x_0\rightarrow -x_0$ and $x\rightarrow -x$.

\subsection{Case with $|x_0|<a$}\label{App_2}
Following the same procedure as in Appendix \ref{x0<-a}, for $|x_0|<a$ the Laplace transform of $P_{\rm nc}(x,t|x_0)$ can be computed, yielding

\begin{figure*}
\includegraphics[width=\textwidth]{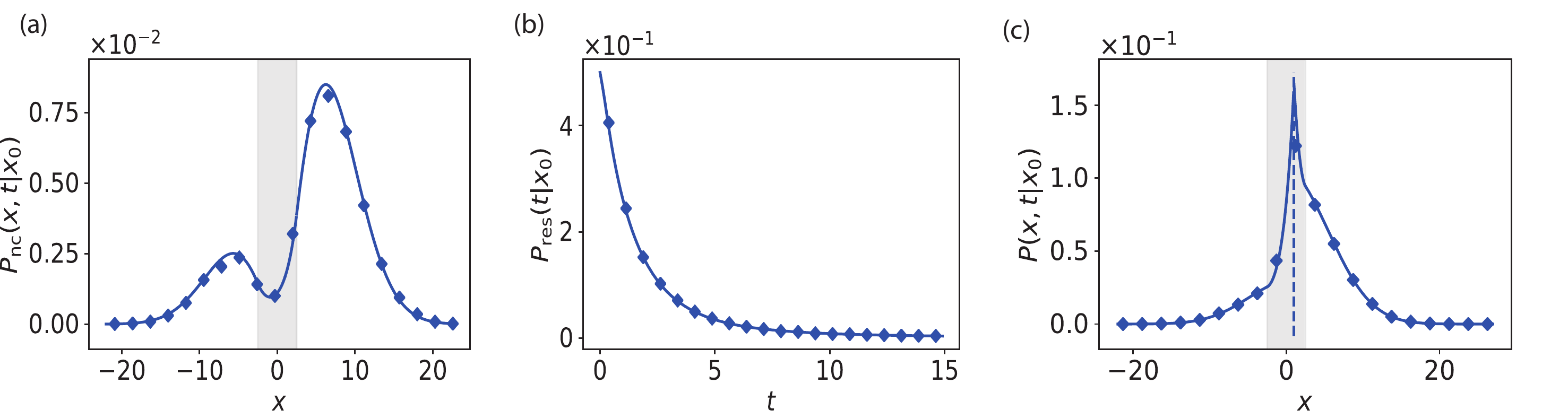}
\caption{\textbf{Statistics of the Brownian tunnelling:} (a)  Probability density $P_{\rm nc}(x,t|x_0)$ at time $t=15$ with $D=1$, $r=0.5$, $a=2.5$, $x_0=1$. The solid line represents the inverse numerical Laplace transform of Eq. \eqref{eq:Pnc2} while the symbols correspond to the result of numerical simulations. The grey shaded area highlights the interval within which resetting occurs. (b) Probability of first reset time $P_{\rm res}(t|x_0)$ up to time $t=15$ with the same parameters as panel (a): comparison between inverse Laplace transform of Eq. \eqref{eq:Pres2} and numerical simulations. (c) Probability density $P(x,t|x_0)$ with $x_r=x_0$, indicated by the vertical dashed line. The solid line represents the inverse Laplace transform of Eq. \eqref{eq:PS2} while symbols indicate the results of numerical simulations. All numerical simulations were done using the Euler numerical integration scheme with time step $\Delta t =0.05$, and they were repeated $N=10^5$ times.} \label{fig:Pnc2}
\end{figure*}

\begin{small}
\begin{equation}\label{eq:Pnc2}
\begin{aligned}
&\tilde{P}_{\rm nc}(x,s|x_0)=\frac{1}{2D}\left\{\left[\mu\sinh\left(a\nu\right)+\nu\cosh\left(a\nu\right)\right]\left[\nu\sinh\left(a\nu\right)+\mu\cosh\left(a\nu\right)\right]\right\}^{-1}\\
&\\[-8pt]
&\begin{cases}
e^{\mu(x+a)}\left[\nu\cosh\left(\nu(a-x_0)\right)+\mu\sinh\left(\nu(a-x_0)\right)\right] &\mbox{for}\,\,x\le -a,\\
&\\[-8pt]
\frac{1}{2\nu}\left[2\mu\nu\sinh\left(\nu(2a+x-x_0)\right)+(\mu^2+\nu^2)\cosh\left(\nu(2a+x-x_0)\right)+\frac{r}{D}\cosh\left(\nu(x_0+x)\right)\right] &\mbox{for}\,\,x\in(-a,x_0),\\
&\\[-8pt]
\frac{1}{2\nu}\left[2\mu\nu\sinh\left(\nu(2a+x_0-x)\right)+(\mu^2+\nu^2)\cosh\left(\nu(2a+x_0-x)\right)+\frac{r}{D}\cosh\left(\nu(x_0+x)\right)\right]&\mbox{for}\,\,x\in[x_0,a],\\
&\\[-8pt]
e^{\mu(a-x)}\left[\nu\cosh\left(\nu(a+x_0)\right)+\mu\sinh\left(\nu(a+x_0)\right)\right]  &\mbox{for}\,\,x>a;
\end{cases}
\end{aligned}
\end{equation}
\end{small}
Once again, as shown in Fig. \ref{fig:Pnc2}a, the resulting $P_{\rm nc}$ as a function of $x$ is exponentially suppressed within the resetting/evaporation area inducing an unbalance of the distribution and the presence of two peaks, whose relative intensity depends on the value of $x_0$.

The Laplace transform of the probability distribution of the first resetting time, combining Eq. \eqref{eq:pres} and Eq. \eqref{eq:Pnc2}, turns out to be

\begin{equation}\label{eq:Pres2}
\tilde{P}_{\rm res}(s|x_0)=\frac{r}{D\nu^2}\left(1-\frac{\mu\cosh(x_0\nu)}{\mu\cosh(a\nu)+\nu\sinh(a\nu)}\right).
\end{equation}
Fig. \ref{fig:Pnc2}b displays the probability $P_{\rm res}(t|x_0)$: being the initial position inside the resetting region, particles are more probable to reset for the first time at short times. 

Finally, plugging Eq. \eqref{eq:Pnc2} and Eq. \eqref{eq:Pres2} in Eq. \eqref{psxr}, $\tilde{P}(x,s|x_0)$ reads
\begin{small}
\begin{equation}
\begin{aligned}
&\tilde{P}(x,s|x_0)=\frac{\nu^2}{2D\mu}\left\{\left[\mu\sinh\left(a\nu\right)+\nu\cosh\left(a\nu\right)\right]\left[\mu\nu\sinh\left(a\nu\right)+\mu^2\cosh\left(a\nu\right)+\frac{r}{D}\cosh\left(x_0 \nu\right)\right]\right\}^{-1}\\
&\\[-8pt]
&\begin{cases}
e^{\mu(x+a)}\left[\nu\cosh\left(\nu(a-x_0)\right)+\mu\sinh\left(\nu(a-x_0)\right)\right] &\mbox{for}\,\,x<-a,\\
&\\[-8pt]
\frac{1}{2\nu}\left[2\mu\nu\sinh\left(\nu(2a+x-x_0)\right)+(\mu^2+\nu^2)\cosh\left(\nu(2a+x-x_0)\right)+\frac{r}{D}\cosh\left(\nu(x_0+x)\right)\right] &\mbox{for}\,\,x\in(-a,x_0),\\
&\\[-8pt]
\frac{1}{2\nu}\left[2\mu\nu\sinh\left(\nu(2a+x_0-x)\right)+(\mu^2+\nu^2)\cosh\left(\nu(2a+x_0-x)\right)+\frac{r}{D}\cosh\left(\nu(x_0+x)\right)\right]&\mbox{for}\,\,x\in(x_0,a),\\
&\\[-8pt]
e^{\mu(a-x)}\left[\nu\cosh\left(\nu(a+x_0)\right)+\mu\sinh\left(\nu(a+x_0)\right)\right]  &\mbox{for}\,\,x>a,
\end{cases}
\end{aligned}\label{eq:PS2}
\end{equation}
\end{small}
and its inverse Laplace transform is reported in Fig. \ref{fig:Pnc2}c. Since the resetting point (vertical dashed line) is inside the resetting region (grey area), particles tend to be more confined with respect to the case with $|x_0|>a$; however, this fact is not sufficient to ensure the existence of a stationary distribution for finite $a$.

\subsection{Moments}\label{a:moments}
In this Section we discuss in more detail the time evolution of the average position $\langle x(t)\rangle$ and variance ${\sigma^2(t)\equiv\langle x^2(t)\rangle-\langle x(t)\rangle^2}$. For simplicity we focus on the case $x_0=x_r$, the results of which can be easily generalized for $x_0\neq x_r.$
As a matter of fact, for $x_r\neq x_0$ the long-time behavior of the moments can be obtained from that of the $x_0=x_r$ case upon substituting $x_0\rightarrow x_r$ : for long times the dynamics depend only on $x_r$. To prove this statement we refer to Eq. \eqref{eq:Pgen2} and we insert the expansion of $\tilde{P}_{\rm res}(s|x_0)=1+O(\sqrt{s})$, see Eq. \eqref{eq:presexp}, to obtain:

\begin{equation}\label{eq:Pexp}
\begin{aligned}
&\tilde{P}(x,s|x_0,x_r)=\tilde{P}(x,s|x_r)+O(\sqrt{s}),
\end{aligned}
\end{equation}
where we have used the fact that $\tilde{P}_{\rm nc}(x,s|x_0)=O(1)$ as $s\rightarrow 0$. From Eq. \eqref{eq:Pexp} we can read explicitly that, at long times, the leading contribution to $P(x,t|x_0,x_r)$ coincides with the same probability distribution with $x_0=x_r.$ It follows that this property holds also for all moments of $P(x,t|x_0,x_r)$.

We do not report here the lengthy expressions of $\langle x(t)\rangle$ and $\sigma^2(t)$ but we focus on their long- and short-time behaviors, which we determine by inverting the leading contribution in the expansion of the Laplace transform for $s\rightarrow 0$ and $s\to\infty$, respectively.

\begin{figure}
\begin{minipage}[b]{7.5cm}
\centering
\includegraphics[width = 3in]{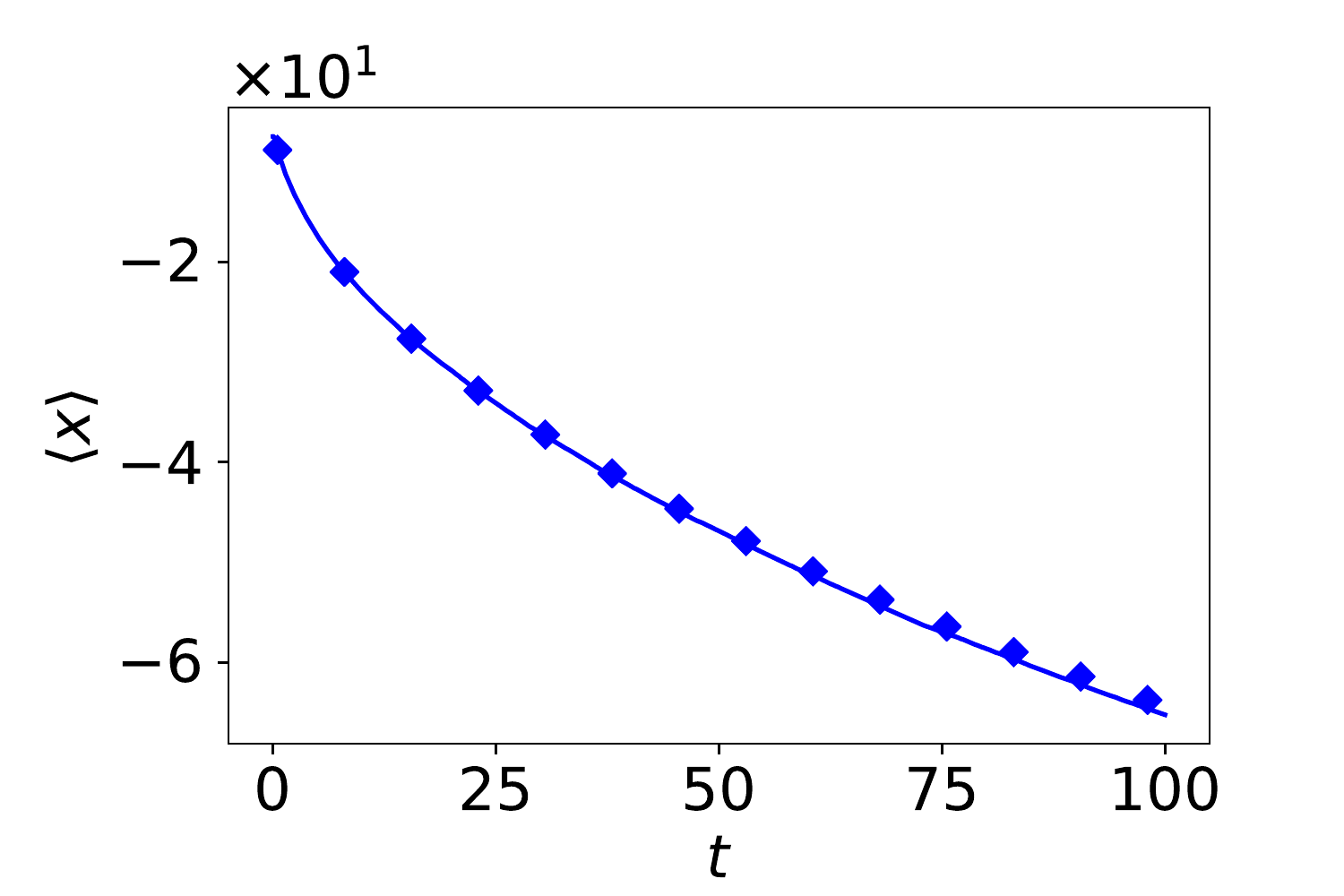}
\caption{\textbf{Brownian tunnelling with resetting:} Comparison between numerical simulation (solid line) and theory (symbols) of the time evolution of the average position $\langle x(t) \rangle$ of the particle  with parameters $D=50$, $a=5$, $r=2.5$ and $x_0=x_r=-7.5$. Simulations are performed with a time step $\Delta t =0.05$, and are repeated $N=10^5$ times.}
\label{av1}
\end{minipage}
\ \hspace{2mm} \hspace{3mm} \
\begin{minipage}[b]{7.5cm}
\centering
\includegraphics[width = 3in]{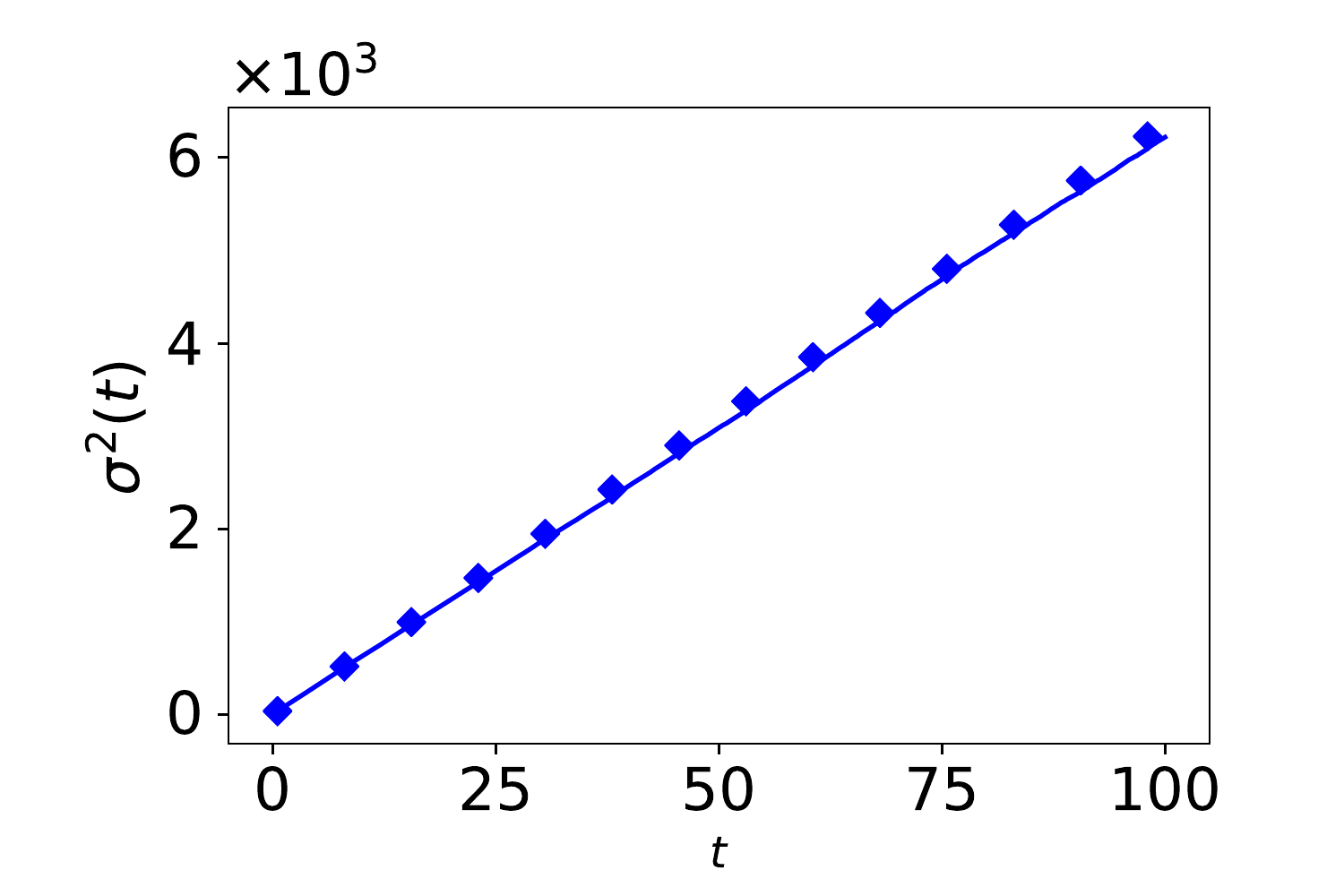}
\caption{\textbf{Brownian tunnelling with resetting:} Comparison between numerical simulation (solid line) and theory (symbols) of the time evolution of the variance $\langle \sigma^2(t) \rangle$ of the particle  with parameters $D=50$, $a=5$, $r=2.5$ and $x_0=x_r=-7.5$. Simulations are performed with a time step $\Delta t =0.05$, and are repeated $N=10^5$ times.}
\label{var1}
\end{minipage}
\end{figure}

First, we analyse $\langle x(t)\rangle$ and $\sigma^2(t)$ resulting from the probability distribution in Eq. \eqref{eq:PS1}, corresponding to $|x_0|>a$.
We identify two regimes for $\langle x(t)\rangle$: at short times, until the particle reaches the resetting region, the average position is constant $\langle
x(t)\rangle=x_0+o(1)$ because of free diffusion; for larger times, instead, the average position grows as $\sqrt{t}$ according to

\begin{equation}\label{av_1_asym}
\begin{aligned}
\langle x(t) \rangle =\sign(y_0)\,2\sqrt{\frac{Dt}{\pi}}\phi(y_0,\rho)+O\left(t^0\right),
\end{aligned}
\end{equation}
where $\phi(y_0,\rho)$ is reported in Eq. \eqref{eq:xfull}. 
Resetting, as suggested by Eq. \eqref{av_1_asym} tends to confine at long times the particle to the half-plane containing the resetting point $x_0$.
Notice that, for $|y_0|>1$, $0<\phi(y_0,\rho)<1$, and, in fact, $\phi$ increases monotonically upon increasing $|y_0|$, attaining its maximum $\phi=1$ at $|y_0|\rightarrow\infty$ and its minimum $\phi=\tanh^2\rho$ at $|y_0|=1.$
All these features are displayed in Figure \ref{av1}, that shows the average position for a particle whose initial position is $x_0<-a$: the solid line stands for simulations while the diamonds for theoretical prediction. 

As in the case of the average position, the variance exhibits, at small times, the Gaussian behavior $\sigma^2(t)= 2D t+o(\sqrt{t})$. 
However, for long times, free diffusion dominates: there are infinite free-diffusive paths corresponding to particles that, because of the recurrence of the one-dimensional Brownian motion, comes back to the resetting region and experience resetting; on long time scales this process happens many times. Accordingly, resetting will tend to localize the motion of the particle, renormalizing the diffusion constant $D$ to a smaller effective value $D_{\rm eff}<D$. Indeed, at long times 
\begin{equation}\label{eq:sigmat}
\sigma^2(t)= 2D_{\rm eff}t+O(\sqrt{t})
\end{equation} 
where the effective diffusion constant, defined as $D_{\rm eff}\equiv\lim_{t\rightarrow\infty}\sigma^2(t)/2D$, is given by Eq. \eqref{eq:defffull}; it is remarkable that $D_{\rm eff}<D$ even for $|y_0|\rightarrow\infty$ still, because recurrence makes the particle feel the presence of the resetting potential, leading to
\begin{equation}
\lim_{|y_0|\rightarrow \infty} \frac{D_{\rm eff}}{D}=1-\frac{2}{\pi}<1.
\end{equation}
This equation provides a lower bound for $D_{\rm eff}$ which is independent of the resetting parameters.
Figure \ref{var1} shows the comparison between simulations of the mean square displacement (solid line) and the theoretical prediction (symbols).

We observe a similar behavior for the motion of a Brownian particle in the presence of a reflecting barrier in the origin at $x=0$, starting from the initial position $x_0$. The probability distribution of its position $x$ reads \cite{jacobs_2010}

\begin{equation}
P_{\rm RB}(x,t|x_0,0)=\frac{e^{-\frac{(x-x_0)^2}{4Dt}}}{\sqrt{D t\pi}\left(1+\erf\left(\frac{|x_0|}{2\sqrt{Dt}}\right)\right)}.
\end{equation}
The mean and the variance of $P_{\rm RB}(x,t|x_0,0)$ are given by

\begin{equation}
\langle x_{\rm RB}(t)\rangle=x_0+\sign(x_0)\,2\sqrt{\frac{Dt}{\pi}}\frac{e^{-\frac{x_0^2}{4Dt}}}{1+\erf\left(\frac{|x_0|}{2\sqrt{Dt}}\right)},
\end{equation}
and
\begin{equation}
\begin{aligned}
\sigma^2_{\rm RB}(t)&=2Dt\left\{1-\frac{2}{\pi}\frac{e^{-\frac{x_0^2}{2Dt}}}{\left[1+\erf\left(\frac{|x_0|}{2\sqrt{Dt}}\right)\right]^2}\right\}-2|x_0|\sqrt{\frac{Dt}{\pi}}\frac{e^{-\frac{x_0^2}{4Dt}}}{1+\erf\left(\frac{|x_0|}{2\sqrt{Dt}}\right)}.
\end{aligned}
\end{equation}
Their expressions in the long-time limit are

\begin{equation}\label{eq:xrb}
\langle x_{\rm RB}(t)\rangle=2\sign(x_0)\sqrt{\frac{Dt}{\pi}}+O(t^0),
\end{equation}
and
\begin{equation}\label{eq:srb}
\sigma^2_{\rm RB}(t)=2Dt\left(1-\frac{2}{\pi}\right)+O(\sqrt{t}),
\end{equation}
which depend only on the sign of $x_0$ and not on its actual value.
This similarity between Eqs. \eqref{av_1_asym} and \eqref{eq:sigmat}  on the one side and Eqs. \eqref{eq:xrb} and \eqref{eq:srb} on the other is not accidental and is explained by the fact that, at long times, the resetting barrier has on average the effect of pushing back particles: this can be visualized as if particles feel a weaker reflecting barrier with an efficiency given by $\phi(y_0,\rho)<1$.
As anticipated, $\phi$ is maximum in the limit $|y_0|\rightarrow\infty$, for which the evolution of the average position (\ref{av_1_asym}) is the same as that of a Brownian particle with reflecting barrier. This suggests the fact the particle behaves as if it gets perfectly reflected and its motion is restricted to an half-plane.

We now consider the case with $|x_0|<a$ (e.g., $|y_0|<1$), corresponding to the probability distribution \eqref{eq:PS2}.
The behavior of the average position and the variance is qualitatively the same as before: at short times it is the same as diffusion until the first resetting event; at long times, instead, it satisfies the same asymptotic expressions as Eqs. \eqref{av_1_asym} and \eqref{eq:sigmat} but with different efficiency, as reported in Eq. \eqref{eq:xfull}.
Even in this case, one has that $0<\phi<1$ and $\phi$ is monotonically increasing upon increasing $|y_0|$: it attains its maximum $\phi=\tanh^2\rho$ at $|y_0|=1$ and its minimum $\phi=0$ at $y_0=0.$ In particular, for $y_0=0$, $\langle x\rangle=0$ at all times by symmetry. 



The variance is always smaller than $2D t$, because even at short times the particle moves in the resetting region. The long-time effective diffusion constant is given once again by Eq. \eqref{eq:defffull}.
The behavior of $D_{\rm eff}$ is shown in Figure \ref{fig:Deff1}b: note that the maximum is attained at ${y_0=0}$, for which $D_{\rm eff}=D$, as a consequence of the fact that at long time, closer $y_0$ to the origin the less the particle will be localized.

\section{RESETTING WITH\\ PERIODIC BOUNDARY CONDITIONS}\label{a:PBC}

Here we address the problem of a diffusing particle which possibly resets when located within a interval with the additional condition that its position is restricted to the segment $x\in(-L, L)$ with periodic boundary condition, i.e., the particle moves along a ring of length $2L$.
As in the case of simple diffusion with periodic boundary conditions and no resetting, the system will show the appearance of a stationary probability distribution \citep{jacobs_2010}. The solution of this problem, as in the case with open boundary, generically depends on the position of the resetting point and of the initial point; for simplicity, in what follows, we set $x_0=x_r$.


\subsection{Case with $-L<x_0<-a$}

The type of solution that we seek for Eq. \eqref{LaplaceSE} is of the form

\begin{equation}\label{eq:PncPBC}
\begin{aligned}
&\tilde{P}_{\rm nc}(x,s|x_0)=\begin{cases}
A_1(s,x_0)\,e^{x\mu}+A_2(s,x_0)\,e^{-x\mu} & \mbox{for}\,\, x\in(-L,x_0),\\
B_1(s,x_0)\,e^{x\mu}+B_2(s,x_0)\,e^{-x\mu} & \mbox{for}\,\, x\in(x_0,a),\\
C_1(s,x_0)\,e^{x\nu}+C_2(s,x_0)\,e^{-x\nu}& \mbox{for}\,\,x\in(-a,a),\\
D_1(s,x_0)\,e^{x\mu}+D_2(s,x_0)\,e^{-x\mu} & \mbox{for}\,\, x\in(a,L),
\end{cases}
\end{aligned}
\end{equation}
with periodic boundary conditions, i.e., $\tilde{P}_{\rm nc}(-L,s|x_0)=\tilde{P}_{\rm nc}(L,s|x_0)$ and $\partial_x\tilde{P}_{\rm nc}(-L,s|x_0)=\partial_x\tilde{P}_{\rm nc}(L,s|x_0)$. These conditions ensure the continuity of the distribution and particle currents.
We fix the eight constants in Eq. \eqref{eq:PncPBC} by requiring the continuity of $\tilde{P}_{\rm nc}(-L,s|x_0)$ in $\{\pm a, x_0, \pm L\}$ (four conditions), the continuity of the first spatial derivative in $\pm a, \pm L$ (three conditions) and the condition \eqref{eq:derjump} in $x_0$.
The solution is given by

\begin{figure*}
\includegraphics[width=\textwidth]{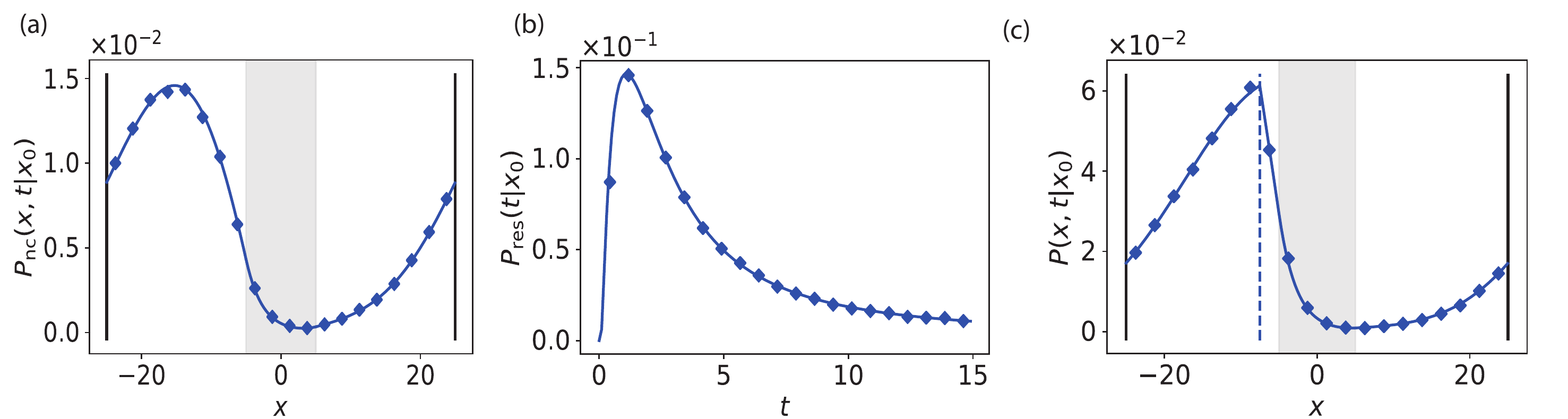}
\caption{\textbf{Statistics of resetting with periodic boundary conditions:} (a)  Probability density $P_{\rm nc}(x,t|x_0)$ at time $t=15$ with $D=5$, $r=1$, $a=5$, $x_0=-7.5$. The solid line represents the inverse numerical Laplace transform of Eq. \eqref{eq:pncPBC1} while symbols correspond to simulations. The grey shaded area indicates the region within which resetting may occur. (b) Probability of first reset time $P_{\rm res}(t|x_0)$ as a function of $t$ with same parameters: comparison between inverse Laplace transform of Eq. \eqref{eq:presPBC1} and simulations. (c) Probability density $P_{\rm nc}(x,t|x_0)$ as a function of position $x$ with $x_r=x_0$, indicated by the vertical dashed line. The solid line represents the inverse Laplace transform of Eq. \eqref{eq:pPBC1} while diamonds simulations. Simulations were done using the Euler numerical integration scheme with time step $\Delta t =0.05$, and they were repeated $N=10^5$ times. }\label{fig:PLnc1}
\end{figure*}

\begin{small}
\begin{equation}\label{eq:pncPBC1}
\begin{aligned}
&\tilde{P}_{\rm nc}(x,s|x_0)=\frac{1}{D}\left\{(\mu^2+\nu^2)\sinh\left(2a\nu\right)\sinh\left(2\mu\left(L-a\right)\right)+2\mu\nu\left[\cosh\left(2a\nu\right)\cosh\left(2\mu\left(L-a\right)\right)-1\right]\right\}^{-1}\\
&\\[-8pt]
&\begin{cases}
\left\{\sinh\left(2a\nu\right)\left[\mu^2\cosh\left(\mu(a+x_0)\right)\cosh\left(\mu(2L-a+x)\right)-\nu^2\sinh\left(\mu(a+x_0)\right)\sinh\left(\mu(2L-a+x)\right)\right]\right.\\
\,\,\,\,\,\,\,\,\,\,\left.+\mu\nu\left[\cosh\left(2a\nu\right)\sinh\left(\mu\left(2L+x-x_0-2a\right)\right)-\sinh(\mu(x-x_0))\right]\right\}/\mu & \mbox{for}\,\, x\in[-L,x_0],\\
&\\[-8pt]
\left\{\sinh\left(2a\nu\right)\left[\mu^2\cosh\left(\mu(a+x)\right)\cosh\left(\mu(2L-a+x_0)\right)-\nu^2\sinh\left(\mu(a+x)\right)\sinh\left(\mu(2L-a+x_0)\right)\right]\right.\\
\,\,\,\,\,\,\,\,\,\,\left.+\mu\nu\left[\cosh\left(2a\nu\right)\sinh\left(\mu\left(2L+x_0-x-2a\right)\right)+\sinh(\mu(x-x_0))\right]\right\}/\mu & \mbox{for}\,\, x\in(x_0,-a),\\
&\\[-8pt]
\mu\left[\cosh\left(\mu(a+x_0)\right)\sinh\left(\nu(a+x)\right)+\cosh(\mu(2L+x_0-a))\sinh\left(\nu(a-x)\right)\right]\\
\,\,\,\,\,\,\,\,\,\,-\nu\left[\sinh\left(\mu(a+x_0)\right)\cosh\left(\nu(a+x)\right)-\sinh(\mu(2L+x_0-a))\cosh\left(\nu(a-x)\right)\right]& \mbox{for}\,\, x\in[-a,a],\\
&\\[-8pt]
\left\{\sinh\left(2a\nu\right)\left[\mu^2\cosh\left(\mu(a+x_0)\right)\cosh\left(\mu(x-a)\right)-\nu^2\sinh\left(\mu(a+x_0)\right)\sinh\left(\mu(x-a)\right)\right]\right.\\
\,\,\,\,\,\,\,\,\,\,\left.+\mu\nu\left[\sinh(\mu(2L+x_0-x))-\cosh\left(2a\nu\right)\sinh\left(\mu\left(2a-x+x_0\right)\right)\right]\right\}/\mu & \mbox{for}\,\, x\in(a,L),\\
\end{cases}
\end{aligned}
\end{equation}
\end{small}
and it can be checked that in the limiting case, $r=0$, of no resetting it reproduces the Laplace transform of the probability density of particles diffusing on the segment $(-L,L)$ with periodic boundary conditions, i.e.,

\begin{equation}
\tilde{P}(x,s|x_0)=\frac{\cosh\left(\mu(L-|x-x_0|)\right)}{2D\mu\sinh(\mu L)},
\end{equation}
whose inverse Laplace transform can be exactly computed as
\begin{equation}\label{eq:diffusionPBC}
P(x,t|x_0,0)=\frac{1}{2L}\left[1+2\sum_{n=1}^\infty(-1)^ne^{-\left(\frac{n\pi}{L}\right)^2D\,t}\cos\left(n\pi\chi\right)\right],
\end{equation}
with $\chi\equiv1-|x-x_0|/L$, see Ref. \cite{Laplace_book}; it is immediate to check that Eq. \eqref{eq:diffusionPBC} has a uniform stationary distribution $P_{\rm st}$ given by $P_{\rm st}(x)=1/2L.$
Figure \ref{fig:PLnc1}a  shows a snapshot of $P_{\rm nc}(x,t|x_0)$ as a function of $x$: the solid line represents the inverse Laplace transform of Eq. \eqref{eq:pncPBC1} while the symbols correspond to the result of numerical simulations; in the resetting area (grey) $P(x,t|x_0)$ is exponentially suppressed compared to the values it has outside it as times goes by, due to resetting particles.

From $\tilde{P}_{\rm nc}(x,s|x_0)$ in Eq. \eqref{eq:pncPBC1} it is possible to obtain the Laplace transform $\tilde P_{\rm res}(s|x_0)$ of the probability $P_{\rm res}(t|x_0)$ that the particle resets for the first time at time $t$:

\begin{equation}\label{eq:presPBC1}
\begin{aligned}
\tilde{P}_{\rm res}&(s|x_0)=\frac{(r/\nu D)\sinh(a\nu)\cosh\left(\mu(L+x_0)\right)}{\nu\sinh(a\nu)\cosh\left(\mu(L-a)\right)+\mu\cosh(a\nu)\sinh\left(\mu(L-a)\right)},
\end{aligned}
\end{equation}
which is represented in Fig. \ref{fig:PLnc1}b.

Plugging Eqs. \eqref{eq:pncPBC1} and \eqref{eq:presPBC1} into Eq. \eqref{psxr}, we derive the Laplace transform for the probability distribution of $P(x,t|x_0)$:
\begin{small}
\begin{equation}\label{eq:pPBC1}
\begin{aligned}
&\tilde{P}(x,s|x_0)=\frac{\nu}{4}\left\{\left[\mu\sinh(a\nu)\cosh\left(\mu(L-a)\right)+\nu\cosh(a\nu)\sinh\left(\mu(L-a)\right)\right]\right.\\
&\left.\left[\nu D(\nu\sinh(a\nu)\cosh\left(\mu(L-a)\right)+\mu\cosh(a\nu)\sinh\left(\mu(L-a)\right))-r\sinh(a\nu)\cosh(\mu(L+x_0))\right]\right\}^{-1}\\
&\\[-8pt]
&\begin{cases}
\left\{\sinh\left(2a\nu\right)\left[\mu^2\cosh\left(\mu(a+x_0)\right)\cosh\left(\mu(2L-a+x)\right)-\nu^2\sinh\left(\mu(a+x_0)\right)\sinh\left(\mu(2L-a+x)\right)\right]\right.\\
\,\,\,\,\,\,\,\,\,\,\left.+\mu\nu\left[\cosh\left(2a\nu\right)\sinh\left(\mu\left(2L+x-x_0-2a\right)\right)-\sinh(\mu(x-x_0))\right]\right\}/\mu & \mbox{for}\,\, x\in[-L,x_0],\\
&\\[-8pt]
\left\{\sinh\left(2a\nu\right)\left[\mu^2\cosh\left(\mu(a+x)\right)\cosh\left(\mu(2L-a+x_0)\right)-\nu^2\sinh\left(\mu(a+x)\right)\sinh\left(\mu(2L-a+x_0)\right)\right]\right.\\
\,\,\,\,\,\,\,\,\,\,\left.+\mu\nu\left[\cosh\left(2a\nu\right)\sinh\left(\mu\left(2L+x_0-x-2a\right)\right)+\sinh(\mu(x-x_0))\right]\right\}/\mu & \mbox{for}\,\, x\in(x_0,-a),\\
&\\[-8pt]
\mu\left[\cosh\left(\mu(a+x_0)\right)\sinh\left(\nu(a+x)\right)+\cosh(\mu(2L+x_0-a))\sinh\left(\nu(a-x)\right)\right]\\
\,\,\,\,\,\,\,\,\,\,-\nu\left[\sinh\left(\mu(a+x_0)\right)\cosh\left(\nu(a+x)\right)-\sinh(\mu(2L+x_0-a))\cosh\left(\nu(a-x)\right)\right]& \mbox{for}\,\, x\in[-a,a],\\
&\\[-8pt]
\left\{\sinh\left(2a\nu\right)\left[\mu^2\cosh\left(\mu(a+x_0)\right)\cosh\left(\mu(x-a)\right)-\nu^2\sinh\left(\mu(a+x_0)\right)\sinh\left(\mu(x-a)\right)\right]\right.\\
\,\,\,\,\,\,\,\,\,\,\left.+\mu\nu\left[\sinh(\mu(2L+x_0-x))-\cosh\left(2a\nu\right)\sinh\left(\mu\left(2a-x+x_0\right)\right)\right]\right\}/\mu & \mbox{for}\,\, x\in(a,L).\\
\end{cases}
\end{aligned}
\end{equation}
\end{small}
The probability density $P(x,t|x_0)$ is represented in Fig. \ref{fig:PLnc1}c: as a function of $x$ it features a cusp in correspondence of the resetting point $x=x_r=x_0$ (indicated by dashed vertical line) and a significant reduction of the probability within the resetting (grey) region. 

Periodic boundary conditions, i.e., the geometry of a ring, allow the existence of a stationary distribution of Eq. \eqref{eq:pPBC1}, which is given by ${P_{\rm st}(x|x_0)=\lim_{s\rightarrow0}s\tilde{P}(x,s|x_0)}$ and therefore by

\begin{small}
\begin{equation}\label{eq:pstPBC1}
\begin{aligned}
&P_{\rm st}(a\,y|a\,y_0)=\frac{\rho}{a}\left\{\left[(\ell-1)\rho\cosh(\rho)+\sinh(\rho)\right]\left[2(\ell-1)\rho\cosh(\rho)+(2+\rho^2(1-2\ell-y_0)(1+y_0))\sinh(\rho)\right]\right\}^{-1}\\
&\\[-8pt]
&\begin{cases}
y\rho\sinh(\rho)\left[\sinh(\rho)-(1+y_0)\rho\cosh(\rho)\right]+\frac{\rho}{2}\left[y_0+(2\ell-2-y_0)\cosh(2\rho)\right]+\frac{1}{2}\left[1+\rho^2(1-2\ell)(1+y_0)\right]\sinh(2\rho) & \mbox{for}\,\, y\in[-\ell,y_0],\\
&\\[-8pt]
-y\rho\sinh(\rho)\left[\sinh(\rho)+(2\ell-1+y_0)\rho\cosh(\rho)\right]+\frac{\rho}{2}\left[(2\ell-2+y_0)\cosh(2\rho)-y_0\right]+\frac{1}{2}\left[1-\rho^2(2\ell+y_0-1)\sinh(2\rho)\right] & \mbox{for}\,\, y\in(y_0,-1),\\
&\\[-8pt]
\cosh(\rho)\left[\sinh(\rho)+(\ell-1)\rho\cosh(\rho)\right]-\rho(\ell+y_0)\sinh(\rho)\sinh(y\rho)& \mbox{for}\,\, y\in[-1,1],\\
&\\[-8pt]
 y\rho\sinh(\rho)\left[\sinh(\rho)-(1+y_0)\rho\cosh(\rho)\right]+\frac{\rho}{2}\left[y_0+2\ell-(2+y_0)\cosh(2\rho)\right]+\frac{1}{2}\left[1+\rho^2(1+y_0)\right]\sinh(2\rho)& \mbox{for}\,\, y\in(1,\ell),\\
\end{cases}
\end{aligned}
\end{equation}
\end{small}
where we have introduced the dimensionless variables $\ell\equiv L/a$, $\rho\equiv a\sqrt{r/D}$, $y\equiv x/a$ and $y_0\equiv x_0/a$.
\subsection{Case with $x_0\in(-a,a)$}

In this case the expression of $\tilde{P}_{\rm nc}(x,s|x_0)$ is given by

\begin{small}
\begin{equation}\label{eq:pncPBC2}
\begin{aligned}
&\tilde{P}_{\rm nc}(x,s|x_0)=\frac{1}{D}\left\{(\mu^2+\nu^2)\sinh\left(2a\nu\right)\sinh\left(2\mu\left(L-a\right)\right)+2\mu\nu\left[\cosh\left(2a\nu\right)\cosh\left(2\mu\left(L-a\right)\right)-1\right]\right\}^{-1}\\
&\\[-8pt]
&\begin{cases}
\mu\left[\cosh\left(\mu(a+x)\right)\sinh\left(\nu(a+x_0)\right)+\cosh(\mu(2L+x-a))\sinh\left(\nu(a-x_0)\right)\right]\\
\,\,\,\,\,\,\,\,\,\,-\nu\left[\sinh\left(\mu(a+x)\right)\cosh\left(\nu(a+x_0)\right)-\sinh(\mu(2L+x-a))\cosh\left(\nu(a-x_0)\right)\right] & \mbox{for}\,\, x\in[-L,-a],\\
&\\[-8pt]
\left\{\sinh\left(2\mu(L-a)\right)\left[\mu^2\sinh\left(\nu(a+x)\right)\sinh\left(\nu(a-x_0)\right)+\nu^2\cosh\left(\nu(a+x)\right)\cosh\left(\nu(a-x_0)\right)\right]\right.\\
\,\,\,\,\,\,\,\,\,\,\left.+\mu\nu\left[\cosh\left(2\mu(L-a)\right)\sinh\left(\nu\left(x+2a-x_0\right)\right)-\sinh(\nu(x-x_0))\right]\right\}/\nu & \mbox{for}\,\, x\in(-a,x_0),\\
&\\[-8pt]
\left\{\sinh\left(2\mu(L-a)\right)\left[\mu^2\sinh\left(\nu(a+x_0)\right)\sinh\left(\nu(a-x)\right)+\nu^2\cosh\left(\nu(a+x_0)\right)\cosh\left(\nu(a-x)\right)\right]\right.\\
\,\,\,\,\,\,\,\,\,\,\left.+\mu\nu\left[\cosh\left(2\mu(L-a)\right)\sinh\left(\nu\left(2a+x_0-x\right)\right)+\sinh(\nu(x-x_0))\right]\right\}/\nu & \mbox{for}\,\, x\in[x_0,a],\\
&\\[-8pt]
\mu\left[\cosh\left(\mu(x-a)\right)\sinh\left(\nu(a-x_0)\right)+\cosh(\mu(2L-x-a))\sinh\left(\nu(a+x_0)\right)\right]\\
\,\,\,\,\,\,\,\,\,\,+\nu\left[\sinh\left(\mu(x-a)\right)\cosh\left(\nu(a-x_0)\right)+\sinh(\mu(2L-x-a))\cosh\left(\nu(a+x_0)\right)\right] & \mbox{for}\,\, x\in(a,L),\\
\end{cases}
\end{aligned}
\end{equation}
\end{small}
the inverse transform of which is reported in Fig. 10a as a function of the time $t$. 
\begin{figure*}
\includegraphics[width=\textwidth]{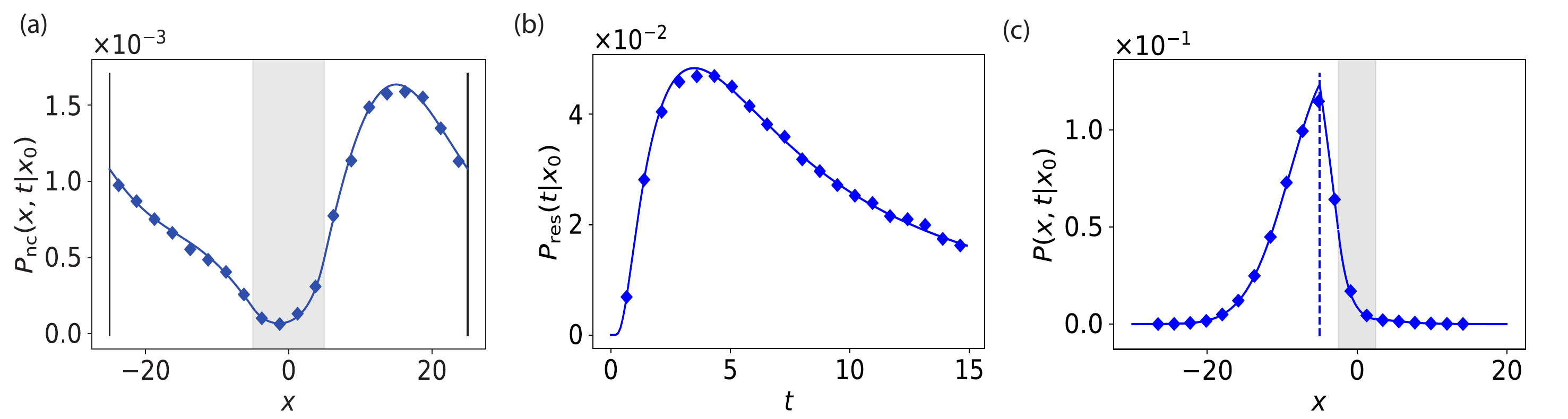}
\caption{\textbf{Statistics of resetting with periodic boundary conditions:} (a)  Probability density $P_{\rm nc}(x,t|x_0)$ at time $t=15$ with $D=5$, $r=1$, $a=5$, $x_0=1.5$. The solid line represents the inverse numerical Laplace transform of Eq. \eqref{eq:pncPBC2} while symbols correspond to simulations. The grey shaded area indicates the region within which resetting may occur. (b) Probability of first reset time $P_{\rm res}(t|x_0)$ as a function of $t$ with same parameters: comparison between inverse Laplace transform of Eq. \eqref{eq:presPBC2} and simulations. (c) Probability density $P(x,t|x_0)$ as a function of position $x$ with $x_r=x_0$, indicated by the vertical dashed line. The solid line represents the inverse Laplace transform of Eq. \eqref{eq:pPBC2} while diamonds simulations. Simulations were done using the Euler numerical integration scheme with time step $\Delta t =0.05$, and they were repeated $N=10^5$ times. }\label{fig:PLnc2}
\end{figure*}

The Laplace transform of the probability $P_{\rm res}(t|x_0)$ that the particle resets for the first time at $t$, computed by plugging Eq. \eqref{eq:Pnc2} in Eq. \eqref{eq:pres}, is

\begin{equation}\label{eq:presPBC2}
\begin{aligned}
\tilde{P}_{\rm res}(s|x_0)=\frac{r}{\nu^2 D}\left[1-\frac{\mu\sinh(a\nu)\cosh\left(\mu(L+x_0)\right)}{\nu\sinh(a\nu)\cosh\left(\mu(L-a)\right)+\mu\cosh(a\nu)\sinh\left(\mu(L-a)\right)}\right],
\end{aligned}
\end{equation}
and that probability, obtained from the inverse transform,  is reported in Fig. \ref{fig:PLnc2}b as a function of time.
From Eqs. \eqref{eq:pncPBC2} and \eqref{eq:presPBC2} one can determine the Laplace transform $\tilde{P}(x,s|x_0)$ of the probability distribution $\tilde P(x,t|x_0)$ according to Eq. \eqref{psxr},

\begin{small}
\begin{equation}\label{eq:pPBC2}
\begin{aligned}
&\tilde{P}(x,s|x_0)=\frac{\nu^2}{4}\left\{\left[\mu\sinh(a\nu)\cosh\left(\mu(L-a)\right)+\nu\cosh(a\nu)\sinh\left(\mu(L-a)\right)\right]\right.\\
&\left.\left[\mu^2 D(\nu\sinh(a\nu)\cosh\left(\mu(L-a)\right)+\mu\cosh(a\nu)\sinh\left(\mu(L-a)\right))-r\mu\cosh(x_0\nu)\sinh(\mu(L-a))\right]\right\}^{-1}\cdot\\
&\\[-8pt]
&\begin{cases}
\mu\left[\cosh\left(\mu(a+x)\right)\sinh\left(\nu(a+x_0)\right)+\cosh(\mu(2L+x-a))\sinh\left(\nu(a-x_0)\right)\right]\\
\,\,\,\,\,\,\,\,\,\,-\nu\left[\sinh\left(\mu(a+x)\right)\cosh\left(\nu(a+x_0)\right)-\sinh(\mu(2L+x-a))\cosh\left(\nu(a-x_0)\right)\right] & \mbox{for}\,\, x\in[-L,-a],\\
&\\[-8pt]
\left\{\sinh\left(2\mu(L-a)\right)\left[\mu^2\sinh\left(\nu(a+x)\right)\sinh\left(\nu(a-x_0)\right)+\nu^2\cosh\left(\nu(a+x)\right)\cosh\left(\nu(a-x_0)\right)\right]\right.\\
\,\,\,\,\,\,\,\,\,\,\left.+\mu\nu\left[\cosh\left(2\mu(L-a)\right)\sinh\left(\nu\left(x+2a-x_0\right)\right)-\sinh(\nu(x-x_0))\right]\right\}/\nu & \mbox{for}\,\, x\in(-a,x_0),\\
&\\[-8pt]
\left\{\sinh\left(2\mu(L-a)\right)\left[\mu^2\sinh\left(\nu(a+x_0)\right)\sinh\left(\nu(a-x)\right)+\nu^2\cosh\left(\nu(a+x_0)\right)\cosh\left(\nu(a-x)\right)\right]\right.\\
\,\,\,\,\,\,\,\,\,\,\left.+\mu\nu\left[\cosh\left(2\mu(L-a)\right)\sinh\left(\nu\left(2a+x_0-x\right)\right)+\sinh(\nu(x-x_0))\right]\right\}/\nu & \mbox{for}\,\, x\in[x_0,a],\\
&\\[-8pt]
\mu\left[\cosh\left(\mu(x-a)\right)\sinh\left(\nu(a-x_0)\right)+\cosh(\mu(2L-x-a))\sinh\left(\nu(a+x_0)\right)\right]\\
\,\,\,\,\,\,\,\,\,\,+\nu\left[\sinh\left(\mu(x-a)\right)\cosh\left(\nu(a-x_0)\right)+\sinh(\mu(2L-x-a))\cosh\left(\nu(a+x_0)\right)\right] & \mbox{for}\,\, x\in(a,L),\\
\end{cases}
\end{aligned}
\end{equation}
\end{small}

whose stationary distribution is given by ${P_{\rm st}(x|x_0)=\lim_{s\rightarrow0}s\tilde{P}(x,s|x_0)}$, which yields

\begin{small}
\begin{equation}\label{eq:pstPBC2}
\begin{aligned}
&P_{\rm st}(a\,y|a\,x_0)=\frac{\rho}{2a}\left\{\left[(\ell-1)\rho\cosh(y_0\rho)+\sinh(\rho)\right]\left[(\ell-1)\rho\cosh(\rho)+\sinh(\rho)\right]\right\}^{-1}\\
&\\[-8pt]
&\begin{cases}
\left[\rho(\ell-1)\cosh(\rho)\cosh(y_0\rho)+\sinh(\rho)(\cosh(y_0\rho)-\rho(\ell+y)\sinh(y_0\rho))\right] & \mbox{for}\,\, y\in[-\ell,-1],\\
&\\[-8pt]
\left[\rho(\ell-1)\cosh((1+y)\rho)\cosh((1-y_0)\rho)+\sinh(\rho)\cosh((1+y-y_0)\rho)\right] & \mbox{for}\,\, y\in(-1,y_0),\\
&\\[-8pt]
\left[\rho(\ell-1)\cosh((1+y_0)\rho)\cosh((1-y)\rho)+\sinh(\rho)\cosh((1-y+y_0)\rho)\right]& \mbox{for}\,\, y\in[y_0,1],\\
&\\[-8pt]
 \left[\rho(\ell-1)\cosh(\rho)\cosh(y_0\rho)+\sinh(\rho)(\cosh(y_0\rho)+\rho(\ell-y)\sinh(y_0\rho))\right] & \mbox{for}\,\, y\in(1,\ell).\\
\end{cases}
\end{aligned}
\end{equation}
\end{small}

At last we mention the particular case $a=L$ of the previous expression, corresponding to resetting in any position of the ring and leading to

\begin{equation}
\tilde{P}(x,s|x_0)=\frac{\nu}{2D\mu^2}\frac{\cosh(\nu(L-|x-x_0|))}{\sinh(L\nu)},
\end{equation}  
which can be analytically inverted to obtain the expression in the time domain:
\begin{equation}\label{eq:PBC_all}
\begin{aligned}
P(x,t|x_0,0)&=\frac{1}{2}\sqrt{\frac{r}{D}}\frac{\cosh(\sqrt{\frac{r}{D}}(L-|x-x_0|))}{\sinh(L\sqrt{\frac{r}{D}})}+\sum_{n=1}^\infty(-1)^n\frac{n^2\pi^2}{L^3}\frac{e^{-Dr_n\,t}}{r_n}\cos\left(n\pi\chi\right),
\end{aligned}
\end{equation}
where $r_n\equiv\frac{n^2\pi^2}{L^2}+\frac{r}{D}$, $\chi\equiv 1-|x-x_0|/L$ and the first term corresponds to the stationary distribution.
\section{Resetting current}\label{App:Current}

The existence of a stationary distribution for the particle position $x$ suggests the emergence of a stationary particle current. So far, we have considered the Brownian particle on a segment with periodic boundary conditions which resets within an interval of length $2a$  to the point $x=x_r$. As anticipated in the main text, one may consider giving  resetting a physical interpretation as making the reset particle traveling across a finite region of space (i.e., from its original position to the point of resetting $x_r$) with infinite velocity in an infinitesimal time interval. 
In a ring, particles can only reset clockwise, counterclockwise or both ways, and therefore fixing the resetting protocol amounts at specifying which particles reset in one or the other way. 
Importantly, note that the only  observable in our problem is the particle position/dynamics and its distribution: as long as resetting is instantaneous, is independent of how resetting occurs. 
Formally, one can integrate the right-hand side of Eq. \eqref{ME} in order to define an effective conserved probability current $J(x,t)$, which obeys the continuity equation $\partial_t P(x,t|x_0)=-\partial_x J(x,t)$, where

\begin{equation}\label{eq:prob_current}
J\equiv J_{\rm diff}+J_{\rm res}
\end{equation}
has a diffusive contribution ${J_{\rm diff}=-D\partial_x P(x,t|x_0)}$ and a resetting contribution given by 
\begin{equation}\label{eq:Jres}
\begin{aligned}
J_{\rm res}(x,t)=J_{\rm res}(-L,t)&+\int_{-L}^x \mathrm{d}y\,r_c(y)P(y,t|x_0)-\theta(x-x_r)\int_{-L}^L \mathrm{d}y\,r(y)P(y,t|x_0).
\end{aligned}
\end{equation}
Here, the freedom in the choosing the value of $J_{\rm res}(-L,t)$ corresponds to the ambiguity in the resetting protocol; as we show below the resetting contribution to this probability current coincides with the physical resetting current according to the picture of resetting presented above.  

The resetting current $J_{\rm res}$ depends on the region of space where particles are reset. 
Accordingly, it is useful to express the contribution to the current $J_{\rm res}$ due to particles that reset in a generic interval $x_1<x<x_2$ at time $t$, i.e.,

\begin{equation}\label{eq:R}
\begin{aligned}
R_a(x_1,x_2)&\equiv\int_{x_1}^{x_2}\mathrm{d}y\,r_c(y)P(y,t|x_0)\\
&=r\int_{x_1}^{x_2}\mathrm{d}y\,\theta(a-|y|)P(y,t|x_0),
\end{aligned}
\end{equation}
which can be naturally split in the contribution $R_a^l$ of the particles resetting  clockwise (leftward) and contribution $R_a^r$ of those resetting counterclockwise (rightward) such that
\begin{equation}
R_a(x_1,x_2)=R_a^l(x_1,x_2)+R_a^r(x_1,x_2);
\end{equation}
the time-dependence in $R_a$ is understood to streamline the notation.
Note that the second and third contributions to $J_{\rm res}$ in the right-hand side of Eq. \eqref{eq:Jres} can be expressed in terms of $R_a(-L,x)$ ad $R_a(-L,L)$.
Also, note that $R_a(x_1,x_2)$ is independent of the choice of the resetting rule while $R_a^{r,\,l}(x_1,x_2)$ is. Indeed, fixing the resetting protocol is equivalent to prescribe the amount of particles that contribute to rightward and leftward $R_a^{\,r,l}$ to $J_{\rm res}$.
Therefore, for a specific protocol, we can express the general expression for $J_{\rm res}$  in terms of $R_a^{\,r,\,l}$, we now address this problem.

\begin{figure}
\includegraphics[scale=0.6]{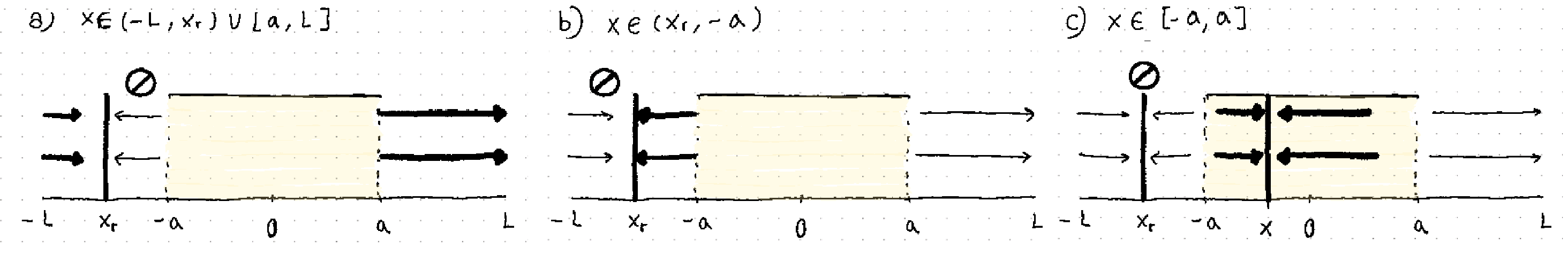}
\caption{Pictorial representation of the contribution for the resetting current $J_{\rm res}(x)$ in the case of $x_r<-a$. The shaded area corresponds to the resetting region in which the resetting current $J_{\rm res}$ originates. Horizontal arrows depict the particles moving towards the resetting point $x_r$ in the two possible directions (left-moving and right-moving): thick arrows represent particle currents that contribute to $J_{\rm res}$ while thin ones to the particles that reach $x_r$ without crossing ("stop" sign) the region in which we compute the current. } 
\label{fig:Current_reset_1}
\end{figure}

Here we first investigate the case in which $x_r\in(-L,-a)$. 
In order to determine the expression of the current $J_{\rm res}$ one has to consider contributions from the different regions of space:
\begin{itemize}
\item[(i)] The current of reset particles through a generic point $x$ within the region $(-L,x_r)$ is only due to right-moving particles which cross the boundary point $L$ and come back from $-L$ to $x_r$ because of periodic boundary conditions, while left-moving ones will stop at $x_r$ without crossing the region under consideration. It may be helpful to refer to Fig. \ref{fig:Current_reset_1}a for a pictorial representation of the current contributions. It follows that the current $J_{\rm res}$ will be positive (right-moving) and equal to

\begin{equation}\label{eq:Jresx1}
J_{\rm res}(x,t)=R_a^r(-a,a).
\end{equation}

\item[(ii)] Analogously, the current through the region $(x_r,-a)$  is only due to left-moving particles, while right-moving ones will cross the boundary $L$ and come back from $-L$ to $x_r$ without crossing the region $(x_r,-a)$, as sketched in Fig. \ref{fig:Current_reset_1}b. This time the current is negative (left-moving) and given by

\begin{equation}\label{eq:Jresx2}
J_{\rm res}(x,t)=-R_a^l(-a,a).
\end{equation}

\item[(iii)] In the region $(-a,a)$, refer to Fig. \ref{fig:Current_reset_1}c, the flux  has contributions coming both from left-moving and right-moving particles whose magnitude depends on the position at which the current is gauged: the left contribution to $J_{\rm res}$ come from the fraction particles $R_a^l$ resetting in $(x,a)$ while the right one $R_a^r$ from the region $(-a,x)$, so that

\begin{equation}\label{eq:Jresx3}
J_{\rm res}(x,t)=R_a^r(-a,x)-R_a^l(x,a).
\end{equation}

\item[(iv)] In the region $(a,L)$ the current is the same as in $(-L,x_r)$ because of periodic boundary conditions.
\end{itemize}

Finally, collecting Eqs. \eqref{eq:Jresx1}, \eqref{eq:Jresx2}  and \eqref{eq:Jresx3}, if the resetting point $x_r$ belongs to the interval $x_r\in(-L,-a)$, the current $J_{\rm res}(x,t)$ across the point $x$ at time $t$ is given by
\begin{equation}\label{eq:J1}
J_{\rm res}(x,t)=
\begin{cases}
 R_a^r(-a,a)& \mbox{for}\,\, x\in[-L,x_r)\cup [a,L],\\
-R_a^l(-a,a) & \mbox{for}\,\, x\in(x_r,-a],\\
R_a^r(-a,x)-R_a^l(x,a)& \mbox{for}\,\, x\in(-a,a).\\
\end{cases}
\end{equation}

Note that the resetting current $J_{\rm res}(x)$ as a function of $x$ is discontinuous in $x_r$ at all times. This is due to the fact that once resetting particles reach $x_r$ they stop. This discontinuity explains the presence of a cusp in the probability density at $x_r$ \cite{evans2011diffusion,Evans_2014}. Moreover, one can easily check that Eq. \eqref{eq:J1} is equivalent to Eq. \eqref{eq:Jres} by identifying $J^p_{\rm res}(-L,t)=R_a^r(-a,a).$
This fact allows to conclude that choosing a resetting protocol is the same as fixing the value of the current at a given point in space. Accordingly, in practice, one may compute once for all $J_{\rm res}(x)-J_{\rm res}(-L)$ which depends only on $P(x,t|x_0)$ and then determine explicitly the value $J_{\rm res}(-L)$ for the resetting protocol adopted in that specific case.
In the same spirit, we express the current for $x_r\in(-a,a)$ as

\begin{figure}
\includegraphics[scale=0.5]{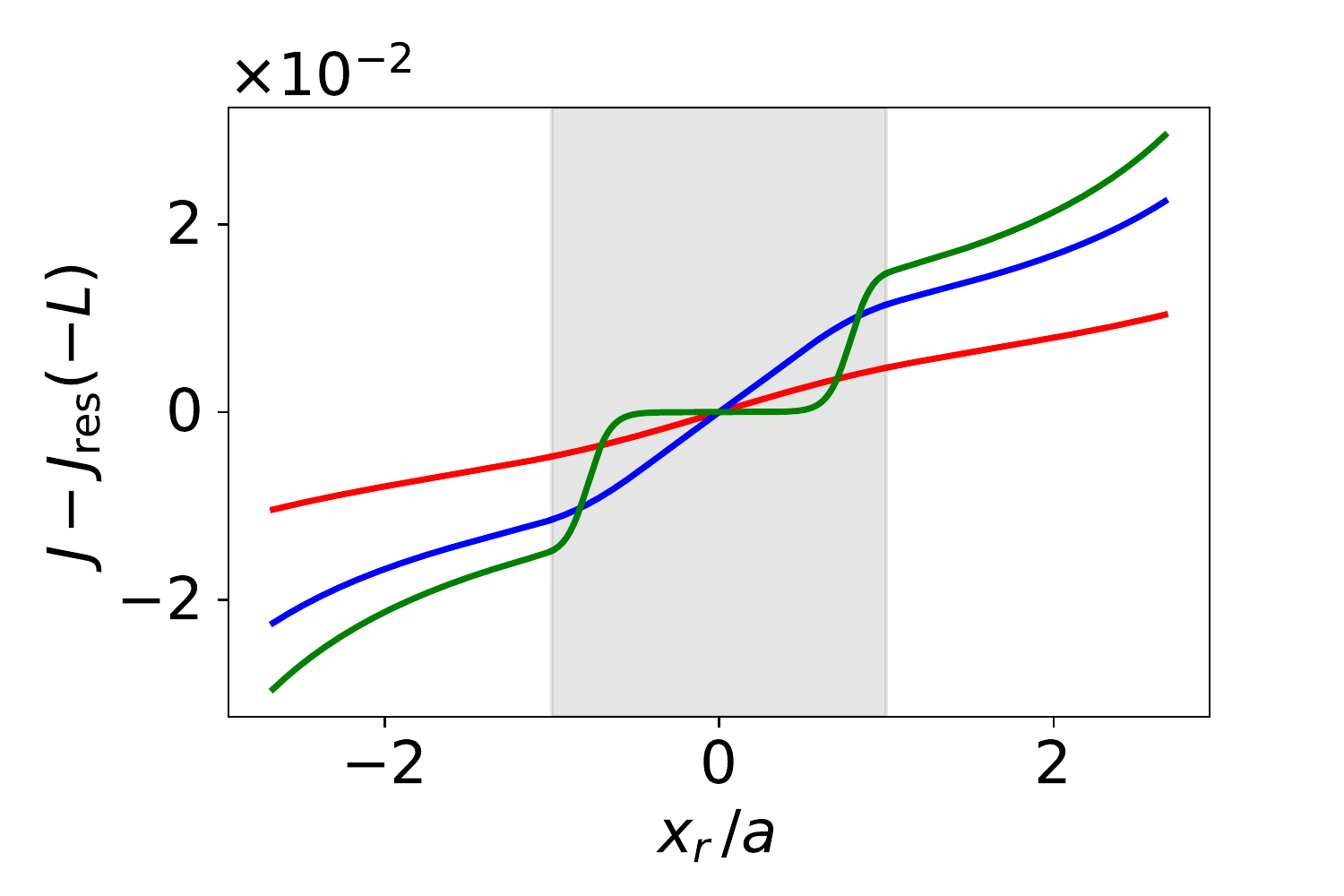}
\caption{Stationary current $J-J_{\rm res}$ in Eq. \eqref{eq:Current0} as a function of the normalized position of the resetting point $x_r/a$ and for various values of $r=0.1$ (red), $1.0$ (blue), $20$ (green) and with $D=5$, $a=7.5$ and 
$L=20$. The grey area corresponds to the resetting region.}\label{fig:Current0}
\end{figure}

\begin{equation}\label{eq:J2}
J_{\rm res}(x,t)=
\begin{cases}
 R_a^r(x_r,a)-R_a^l(-a,x_r)& \mbox{for}\,\, |x|\in [a,L],\\
R_a^r(x_r,x)+R_a^r(x_r,a)-R_a^l(x,x_r) & \mbox{for}\,\, x\in(-a,x_r),\\
R_a^r(x_r,x)-R_a^l(x,a)-R_a^l(-a,x_r)& \mbox{for}\,\, x\in(x_r,a),\\
\end{cases}
\end{equation}
while for $a<x_r<L$

\begin{equation}\label{eq:J3}
\begin{aligned}
&J_{\rm res}(x,t)=\begin{cases}
 -R_a^l(-a,a)& \mbox{for}\,\, x\in[-L,-a]\cup (x_r,L],\\
R_a^r(-a,x)-R_a^l(x,a)& \mbox{for}\,\, x\in (-a,a),\\
R_a^r(-a,a) & \mbox{for}\,\, x\in[a,x_r),\\
\end{cases}
\end{aligned}
\end{equation}
that both satisfy Eq. \eqref{eq:Jres}.
As a last remark, $J_{\rm res}$ for $x_r<-a$ can be obtained from the value of $J_{\rm res}$ for $x_r>a$ by making the following change of variables: $x\leftrightarrow -x$, $x_r\leftrightarrow -x_r$ and $l\leftrightarrow r.$

In the stationary state the total current $J$ in Eq. \eqref{eq:prob_current} is independent of time and space: accordingly, the expression 

\begin{equation}\label{eq:j-jL}
J-J_{\rm res}(-L)=J_{\rm diff}(x)+(J_{\rm res}(x)-J_{\rm res}(-L))
\end{equation}
in the stationary state is constant and independent of the resetting protocol. Therefore, we can reconstruct the value of the total current $J$ by computing $J-J_{\rm res}(-L)$ according to Eq. \eqref{eq:prob_current} and Eq. \eqref{eq:Jres} and then add $J_{\rm res}(-L)$ for the specific protocol.
Note that in the stationary state $J_{\rm diff}(x)+J_{\rm res}(x)$ is also independent of $x$. The expression of $J-J_{\rm res}(-L)$, computed by plugging Eqs. \eqref{eq:pstPBC1} and \eqref{eq:pstPBC2} in Eqs.  \eqref{eq:prob_current} and \eqref{eq:Jres}, as a function of the resetting point is
\begin{equation}\label{eq:Current0}
\frac{J-J_{\rm res}(-L)}{r}=
\begin{cases}
-\frac{\sinh\rho[\sinh\rho-\rho(1+y_r)\cosh\rho]}{\left[(\ell-1)\rho\cosh\rho+\sinh\rho\right]\left[2(\ell-1)\rho\cosh\rho+(2+\rho^2(1-2\ell-y_r)(1+y_r))\sinh\rho\right]}& \mbox{for}\,\, y_r\in[-\ell,-1),\\
\\[-8pt]
\frac{\sinh\rho\sinh(y_r\rho)}{2\left[(\ell-1)\rho\cosh(y_r\rho)+\sinh\rho\right]\left[(\ell-1)\rho\cosh\rho+\sinh\rho\right]}& \mbox{for}\,\, y_r\in[-1,1],\\
\\[-8pt]
\frac{\sinh\rho[\sinh\rho-\rho(1-y_r)\cosh\rho]}{\left[(\ell-1)\rho\cosh\rho+\sinh\rho\right]\left[2(\ell-1)\rho\cosh\rho+(2+\rho^2(1-2\ell+y_r)(1-y_r))\sinh\rho\right]}& \mbox{for}\,\, y_r\in(1,\ell],
\end{cases}
\end{equation}
where $y_r=x_r/a$ and $\ell\equiv L/a$. Eq. \eqref{eq:Current0} can also be seen as the stationary current corresponding to $J_{\rm res}(\pm L)=0$, i.e., the resetting protocol which assumes particles to reset without crossing the points $x=\pm L$.
In Fig. \ref{fig:Current0} we plot the stationary current $(J-J_{\rm res}(-L))/r$  in Eq. \eqref{eq:Current0} as a function of $x_r$ for various values of $r$. Upon increasing $r$, if $x_r$ is inside the resetting region, the current  decreases in magnitude while the opposite happens if $x_r$ is sufficiently outside the resetting region.

In the following Sections, we make two examples of specific protocol choice for which we explicitly compute $J_{\rm res}(-L)$ in order to compute the total stationary current $J$ through Eqs. \eqref{eq:j-jL} and \eqref{eq:Current0}. 

\subsection{Minimal path protocol}

As first example, we consider the \textit{minimal path} protocol: once the particle resets it reaches $x_r$ along the trajectory that minimizes the distance between its current location $x$ and $x_r$, i.e., that of length $\min\left(|x-x_r|,2L-|x-x_r|\right)$. Let us indicate by $l_1$ the path of length $|x_r-x|$ and $l_2$ the path of length $2L-|x-x_r|$. Our aim is to evaluate the value of the current $J_{\rm res}(-L)$. This resetting protocol naturally distinguishes two regions that can contribute to the current:
\begin{itemize}
\item[(i)] First, we consider contribution to $J_{\rm res}(-L)$ which come from the particles resetting in the region $(x_r,L)$. Under this hypothesis, $l_1$ satisfies the minimal path condition if the resetting  particles come from the region $(x_r,L+x_r)$ while $l_2$ does from $(L+x_r,L)$ ($x_r$ can be negative).
In order to understand how particles reset to the left or the right it is necessary to study the problem by distinguishing the possible choices of $x_r$.

If $x_r\in(-L,-a)$, see Fig.\ref{fig:Current_min1}a, the contribution to the current in $\pm L$ is given by the particles that reset to the right since all those that reset to the left stop at $x_r$ before. For this specific case, particles that follow $l_1$ are left-moving while those following $l_2$ are right-moving, and therefore the current $J_{\rm res}(-L)$ we are considering here receives a contribution only from the latter. Accordingly, the value of the current $J_{\rm res}(-L)=R_a(x_r+L,a)$ will be  associated to $l_2$; note that from Eq. \eqref{eq:R} $J_{\rm res}(-L)$ vanishes if $x_r+L\ge a$.

Also for $x_r\in(-a,a)$, see Fig.\ref{fig:Current_min1}b, particles that move to the left are those that reset along $l_1$ and do not contribute to the current. The right-moving resetting particles, associated to $l_2$, contribute by a term $R_a(x_r,L).$

If $x_r>a$ there is no contribution to the current.

\begin{figure}
\begin{minipage}[b]{7.5cm}
\centering
\includegraphics[scale=0.4]{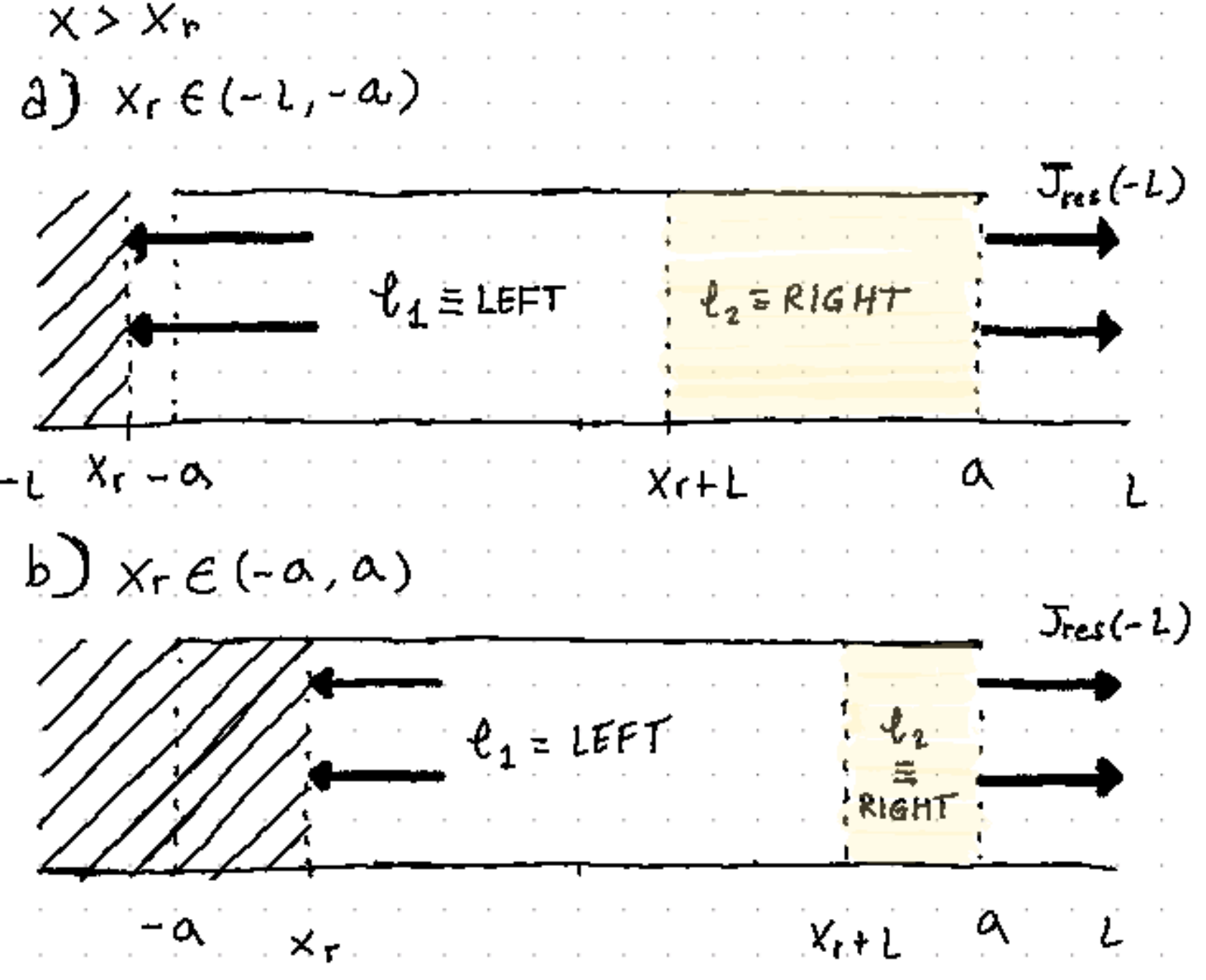}
\caption{Schematic representation of the contribution to the current $J_{\rm res}(-L)$ from particles resetting in the region  $(x_r,L)$ according to the minimal path protocol, the forbidden complementary region $(-L,x_r)$ is denoted by oblique lines. In both cases $x_r\in(-L,-a)$, panel (a), and $x_r\in(-a,a)$, panel (b), the protocol identifies two distinct regions: the yellow shaded region $(x_r+L,a)$ refers to the particles resetting following $l_2$ (right-moving) and the white region $(-a,x_r+L)$ following $l_1$ (left-moving). The resetting particles, horizontal arrows, that contribute to the current $J_{\rm res}(-L)$ are only those following $l_2$. } \label{fig:Current_min1}
\end{minipage}
\ \hspace{2mm} \hspace{3mm} \
\begin{minipage}[b]{7.5cm}
\centering
\includegraphics[scale=0.4]{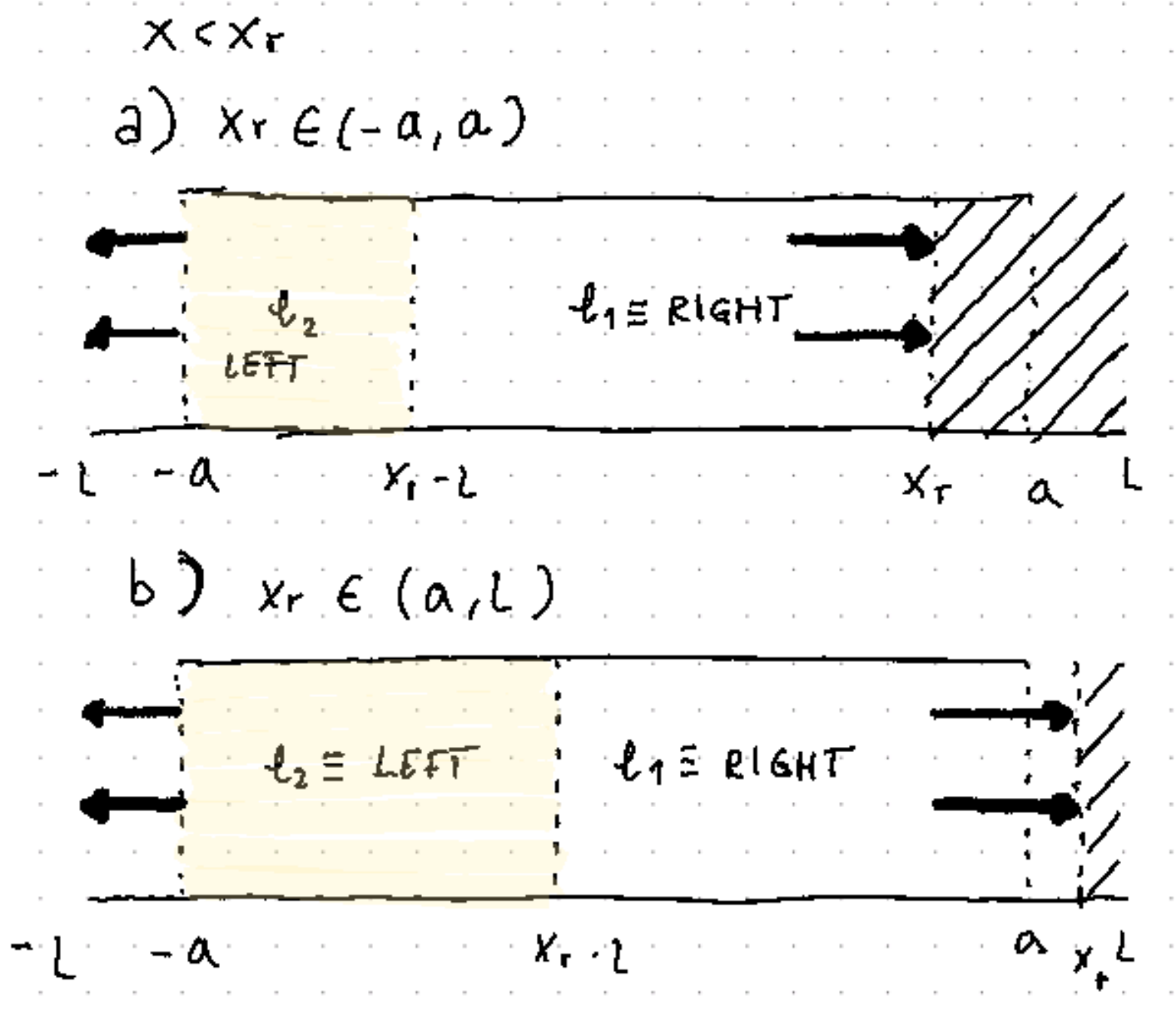}
\caption{Schematic representation of the contribution to the current $J_{\rm res}(-L)$ from particles resetting in the region  $(-L,x_r)$ according to the minimal path protocol, the forbidden complementary region $(x_r,L)$ is denoted by oblique lines. In both cases $x_r\in(-a,a)$, panel (a), and $x_r\in(a,L)$, panel (b), the protocol identifies two distinct regions: the yellow shaded region $(-a,x_r-L)$ refers to the particles resetting following $l_2$ (left-moving) and the white region $(x_r-L,a)$ following $l_1$ (right-moving). The resetting particles, horizontal arrows, that contribute to the current $J_{\rm res}(-L)$ are only those following $l_2$. } \label{fig:Current_min2}
\end{minipage}
\end{figure}

\item[(ii)] If we consider the contribution to $J_{\rm res}(-L)$ due to the particles resetting in $(-L,x_r)$ the path $l_1$ satisfies the minimal path condition within the interval $(x_r-L,x_r)$ while $l_2$ within $(-L,x_r-L)$.

If $x_r\in(-L,-a)$ there is no contribution to the current since $x_r<x_r-L.$

If $x_r\in(-a,a)$, see Fig. \ref{fig:Current_min2}a, we are interested in left-moving particles through $l_2$ resetting in the region $(-a, x_r-L)$, contributing to the current with $-R_a(-a,x_r-L)$.

If $x_r\in(a,L)$, see Fig. \ref{fig:Current_min2}b, the same scenario applies and the contribution to current is $-R_a(-a,x_r-L)$.

\end{itemize} 
Finally, taking into account all the contributions discussed above, the value of the resetting current at $-L$ is eventually 

\begin{equation}\label{eq:JLmin}
\begin{aligned}
&J_{\rm res}(-L,t)=\begin{cases}
 R_a(x_r+L,a)& \mbox{for}\,\, x_r\in(-L,-a),\\
 R_a(x_r+L,a)-R_a(-a,x_r-L)& \mbox{for}\,\, x_r\in[-a,a],\\
-R_a(-a,x_r-L)& \mbox{for}\,\, x_r\in(a,L).
\end{cases}
\end{aligned}
\end{equation}

\begin{figure}
\includegraphics[scale=0.5]{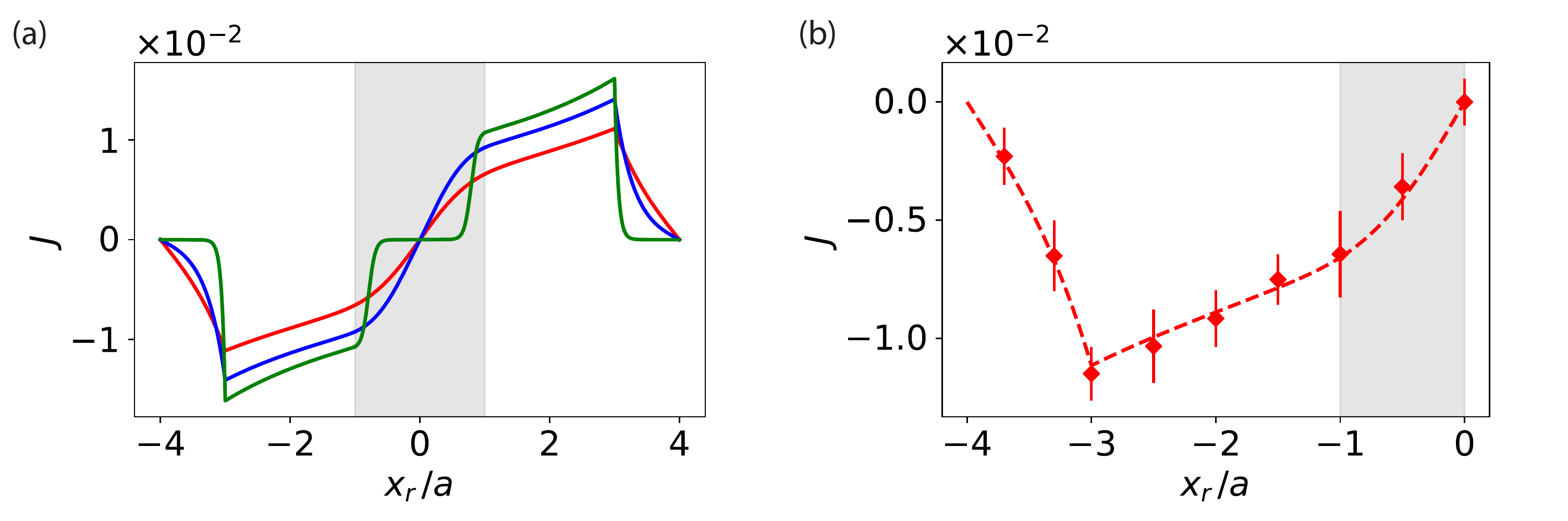}
\caption{(a) Total stationary current $J=J_{\rm diff}+J_{\rm res}$ as a function of $x_r/a$ and for various of $r=0.5$ (red), $2.5$ (blue), $75$ (green) and with $D=5$, $a=5$ and 
$L=20$ according the minimal path protocol.
(b) Comparison between the analytic prediction for $J$ (red curve in (a))  and the numerical simulations (red symbols with error bars). The grey shaded area corresponds to the resetting region. Simulations were done using the Euler numerical integration scheme with time step $\Delta t =0.5$, and they were repeated $N=10^5$ times.}\label{fig:Current_short}
\end{figure}
As already discussed above, the expression of probability distribution depends on the choice of the parameter $x_r$ and so will be the case for $R_a.$ In the stationary state, by considering Eq. \eqref{eq:pstPBC1}, for  $x_r<-a$ one has $R_a(x_r+L,a)$
\begin{equation}
\frac{R_a(a(y_r+\ell),a)}{r}=
\begin{cases}
 0& \mbox{for}\,\, y_r+\ell>1,\\
 \\[-8pt]
 \frac{\left[\sinh\rho-\sinh(\rho(y_r+\ell))\right]\left[\sinh\rho+\rho(\ell-1)\cosh\rho\right]-\rho\sinh\rho(\ell+y_r)\left[\cosh\rho-\cosh(\rho(\ell+y_r))\right]}{\left[(\ell-1)\rho\cosh\rho+\sinh\rho\right]\left[2(\ell-1)\rho\cosh\rho+(2+\rho^2(1-2\ell-y_r)(1+y_r))\sinh\rho\right]}& \mbox{for}\,\, |y_r+\ell|\le 1,\\
 \\[-8pt]
 \frac{2\sinh\rho}{2(\ell-1)\rho\cosh\rho+(2+\rho^2(1-2\ell-y_r)(1+y_r))\sinh\rho}& \mbox{for}\,\, y_r+\ell<-1,
\end{cases}
\end{equation}
while if $x_r\in(-a,a)$, from Eq. \eqref{eq:pstPBC2} one has

\begin{equation}
\frac{R_a(a(y_r+\ell),a)}{r}=
\begin{cases}
 0& \mbox{for}\,\, y_r+\ell>1,\\
 \frac{\rho(\ell-1)\cosh(\rho(1+y_r))\sinh(\rho(1-\ell-y_r))-\sinh\rho\left[\sinh(y_r\rho)+\sinh(\rho(\ell-1))\right]}{2\left[(\ell-1)\rho\cosh(y_r\rho)+\sinh\rho\right]\left[(\ell-1)\rho\cosh\rho+\sinh\rho\right]}& \mbox{for}\,\, y_r+\ell\le 1;
\end{cases}
\end{equation}
similarly for $R_a(-a,x_r-L)$ one has, for $x_r\in(-a,a)$
\begin{equation}
\frac{R_a(-a,a(y_r-\ell))}{r}=
\begin{cases}
 0& \mbox{for}\,\, y_r-\ell<-1,\\
 \frac{\rho(\ell-1)\cosh(\rho(1-y_r))\sinh(\rho(1-\ell+y_r))+\sinh\rho\left[\sinh(y_r\rho)-\sinh(\rho(\ell-1))\right]}{2\left[(\ell-1)\rho\cosh(y_r\rho)+\sinh\rho\right]\left[(\ell-1)\rho\cosh\rho+\sinh\rho\right]}& \mbox{for}\,\, y_r-\ell\ge -1;
\end{cases}
\end{equation}
while, for $x_r>a$,

\begin{equation}
\frac{R_a(-a,a(y_r-\ell))}{r}=
\begin{cases}
 0& \mbox{for}\,\, y_r-\ell<-1,\\
 \\[-8pt]
 -\frac{\left[\sinh\rho-\sinh(\rho(-y_r+\ell))\right]\left[\sinh\rho+\rho(\ell-1)\cosh\rho\right]-\rho(\ell-y_r)\left[\cosh\rho-\cosh(\rho(\ell-y_r))\right]}{\left[(\ell-1)\rho\cosh\rho+\sinh\rho\right]\left[2(\ell-1)\rho\cosh\rho+(2+\rho^2(1-2\ell+y_r)(1-y_r))\sinh\rho\right]}& \mbox{for}\,\, |y_r-\ell|\le 1,\\
 \\[-8pt]
- \frac{2\sinh\rho}{2(\ell-1)\rho\cosh\rho+\left[2+\rho^2(1-2\ell+y_r)(1-y_r)\right]\sinh\rho}& \mbox{for}\,\, y_r-\ell>1.
\end{cases}
\end{equation}
Figure \ref{fig:Current_short} shows the comparison between analytical calculations and numerical simulations for the total stationary current $J$ as a function of the resetting point $x_r$ for a fixed resetting interval and various values of the parameters. The cusp in $|x_r|=L-a$ is due to the vanishing of $J_{\rm res}(-L)$ in Eq. \eqref{eq:JLmin} for $a<|x_r|<L-a$, i.e., no particles resets along $l_2$ and the total stationary current $J$ is only given by Eq. \eqref{eq:Current0}; for $L-a<|x_r|<L$ Eq. \eqref{eq:JLmin} contributes inducing the cusp.
We conclude that, given that $J-J_{\rm res}(-L)$ in Eq. \eqref{eq:Current0} increases monotonously with $x_r$, and that adding $J_{\rm res}(-L)$ reduces the magnitude of $J$ in $|x_r|=L-a$, these cusp points are also maxima for $J$. 

\subsection{Constant rate protocol}

As a second example, not reported in the main text, we consider the case in which a particle that resets has a probability $0<\lambda\le 1$ to follow a rightward resetting direction or probability $1-\lambda$ to reset leftward.
This translates immediately into

\begin{equation}
\begin{aligned}
R_a^r(x_1,x_2)&=\lambda R_a(x_1,x_2)\\
R_a^l(x_1,x_2)&=(1-\lambda) R_a(x_1,x_2).
\end{aligned}
\end{equation}

\begin{figure}
\includegraphics[scale=0.5]{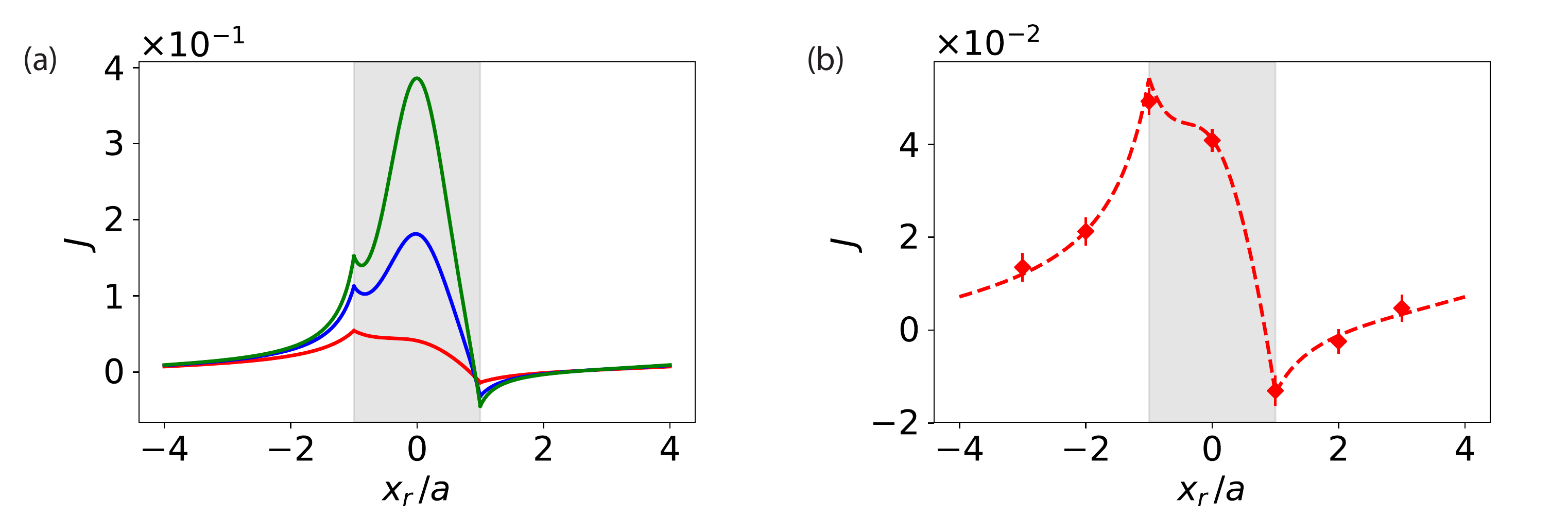}
\caption{(a) Total stationary current $J=J_{\rm diff}+J_{\rm res}$ as a function of $x_r/a$ and different values of $r=0.5$ (red), $1.5$ (blue), $2.5$ (green) and with $D=5$, $a=5$, $L=20$ and $\lambda=0.75$ fixed.
(b) Comparison between the analytic prediction for $J$ (red curve in (a))  and the numerical simulations (red symbols with error bars). The grey shaded area corresponds to the resetting region. Simulations were done using the Euler numerical integration scheme with time step $\Delta t =0.2$, and they were repeated $N=10^5$ times.}\label{fig:Current_lambda_1}
\end{figure}
The current will be fixed by the value of $J_{\rm res}(-L)$, whose expression is given by

\begin{equation}\label{eq:jreflambda}
\begin{aligned}
&J_{\rm res}(-L,t)=\begin{cases}
\lambda R_a(-a,a)& \mbox{for}\,\, x_r\in(-L,-a),\\
\lambda R_a(x_r,a)-(1-\lambda)R_a(-a,x_r)& \mbox{for}\,\, x_r\in(-a,a),\\
-(1-\lambda) R_a(-a,a)& \mbox{for}\,\, x_r\in(a,L),
\end{cases}
\end{aligned}
\end{equation}
where, in the stationary limit, from Eq. \eqref{eq:pstPBC1} one has for $|x_r|>a$ ($|y_r|>1$)
\begin{equation}
R_a(-a,a)/r=\frac{2\sinh\rho}{2(\ell-1)\rho\cosh\rho+[2+\rho^2(1-2\ell+|y_r|)(1-|y_r|)]\sinh\rho},
\end{equation}
while for $|x_r|<a$ ($|y_r|<1$) from \eqref{eq:pstPBC2} one has
\begin{equation}\label{eq:ra2}
\begin{aligned}
&R_a(-a,a\,y_r)/r= \frac{\rho(\ell-1)\cosh(\rho(1-y_r))\sinh(\rho(1+y_r))+\sinh\rho[\sinh(\rho)+\sinh(y_r\rho)]}{2\left[(\ell-1)\rho\cosh(y_r\rho)+\sinh\rho\right]\left[(\ell-1)\rho\cosh\rho+\sinh\rho\right]},\\
&\\[-8pt]
&R_a(a\,y_r,a)/r= \frac{\rho(\ell-1)\cosh(\rho(1+y_r))\sinh(\rho(1-y_r))-\sinh\rho[\sinh(y_r\rho)-\sinh(\rho)]}{2\left[(\ell-1)\rho\cosh(y_r\rho)+\sinh\rho\right]\left[(\ell-1)\rho\cosh\rho+\sinh\rho\right]}.
\end{aligned}
\end{equation}

In Fig. \ref{fig:Current_lambda_1} the total stationary current $J$ is plotted as function of the resetting point $x_r$. For the specific choice, being $\lambda>1/2$ particles reset mostly to the right, making the current positive for almost all values of $x_r$. The current $J$ within the region $|x_r|>a$ increases monotonously as a function of $x_r$ because it is the sum of the two monotonously increasing functions $J_{\rm res}(-L)$ in Eq. \eqref{eq:jreflambda}  and $J-J_{\rm res}(-L)$ in Eq. \eqref{eq:Current0}. Inside the resetting region (grey area) the current $J$ grows with $r$ and, for $r$ sufficiently large, it attains its maximum at $x_r=0$. This feature follows from the competition of two mechanisms. The first consists in the fact that, as $x_r$ becomes larger, more particles reset and contribute negatively to $J_{\rm res}(-L)$ in Eq. \eqref{eq:jreflambda} , leading, for small $r$, to the decrease of the total current. On the other hand, this mechanism becomes less effective as $r$ grows, because the particles reset more and tend to concentrate around $x_r$. Therefore, even for $\lambda-1/2$ slightly positive (negative), the positive (negative) contribution to the current $J_{\rm res}(-L)$ prevails on the negative (positive) one. In the particular case of $x_r=0$, the fraction of resetting particles $R_a(-a,a)$ increases with $r$ and  the two regions $(-a,0)$ and $(0,a)$ contribute equally, i.e., $R_a(-a,0)=R_a(0,a)$ in Eq. \eqref{eq:ra2}, hence  $J=J_{\rm res}(-L)=(\lambda-1/2)R_a(-a,a)$: its magnitude is maximal when $|\lambda-1/2|=1/2$ (completely asymmetric resetting) and minimal at $\lambda=1/2$ (completely symmetric).

\section{APPLICATION TO RNA POLYMERASE}\label{App:Pol}

In this Section, we report the analytical study of model describing the RNA polymerase backtracking. We address the  problem of the motion of the Brownian particle starting in $x_0\in(0,a)$ at a rate $r$ which may reset to the origin when moving within the region $(0,a)$ with the addition of an absorbing boundary in $x_a=x_r=0.$ 
We remark that previous to our work, only analytical solutions for the case $a=\infty$ have been derived in Ref.~\cite{roldan2016stochastic}. 

\begin{figure}\label{fig:S}
\includegraphics[scale=0.5]{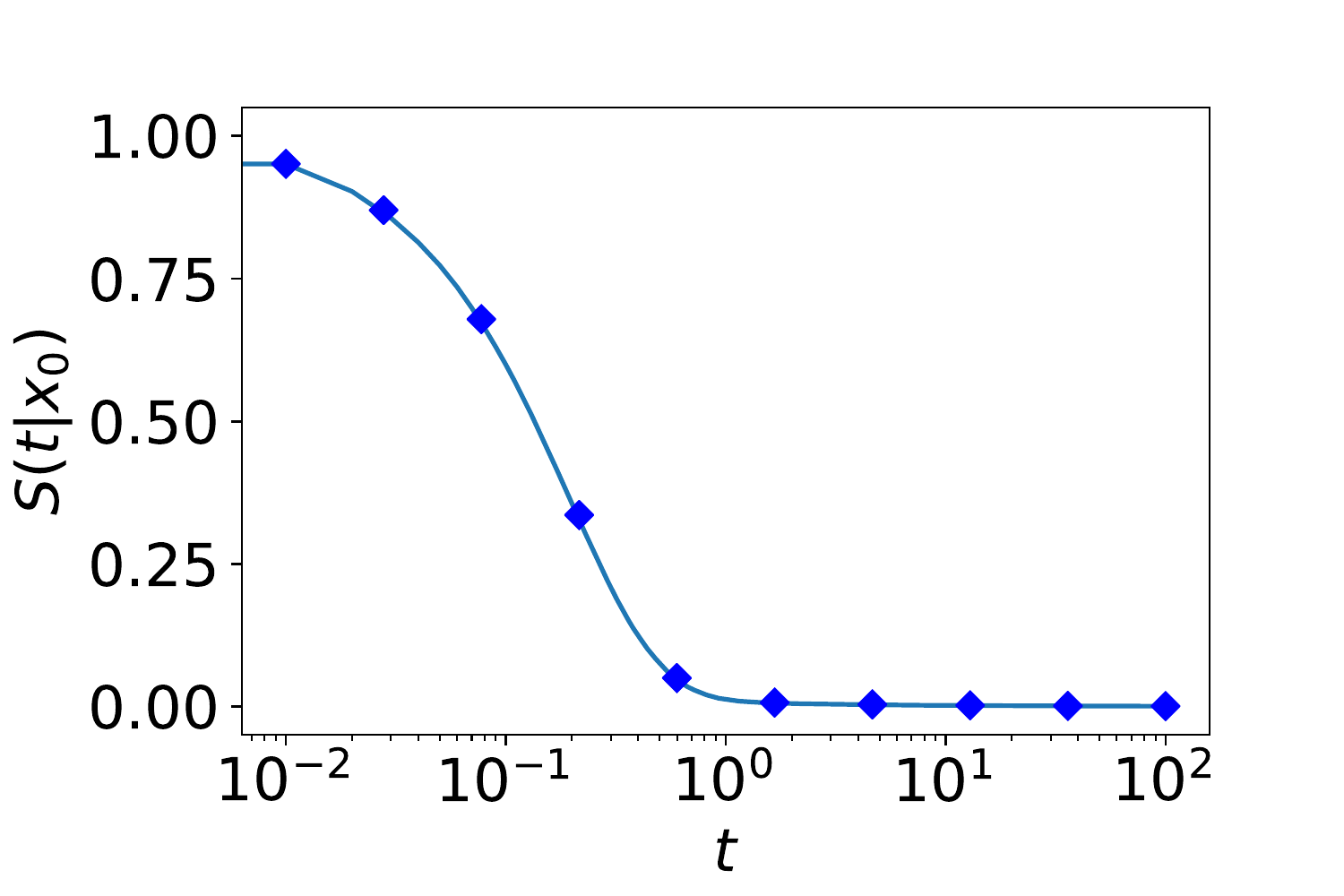}
\caption{Survival probability $S(t)$ with parameters $r=1$, $D=10$, $a=10$ and $x_0=5.$ Solid line indicates the result of numerical simulations and indicates the result of numerical inverse Laplace transform of Eq. \eqref{eq:SL}. Simulations were done using the Euler numerical integration scheme with time step $\Delta t =0.01$, and they were repeated $N=10^5$ times.}\label{Fig:Survival} 
\end{figure}

In this model, the absorbing boundary condition in the origin translates in the vanishing of $P_a(x,t|x_0)$, the probability density of the particle position $x$ at time $t$ starting at $x_0$, in the absorption point $x_a=0$.
Thanks to the linearity of Eq. \eqref{ME} one can obtain $P_a(x,t|x_0)$ from $P(x,t|x_0)$ by means of the method of image charges, e.g., see \citep{redner_2001}. This is accomplished by positioning  negatively ``charged" source of particles originating from $-x_0$ which reset in $(-a,0)$ to the same $x_r=0$, whose distribution is $P(x,t|-x_0)$, yielding

\begin{equation}
P_a(x,t|x_0)=P(x,t|x_0)-P(x,t|-x_0).
\end{equation}
The Laplace transform of this expression is readily computed by considering the expression of $\tilde{P}$ in Eq. \eqref{psxr} in terms of the Laplace transform probability density $P_{\rm nc}(x,t|x_0)$, the probability density for the particles not to reset in the time interval $(0,t)$ moving from $x_0$ to $x$, and $P_{\rm res}(t|x_0)$, the probability distribution of the first resetting time (Appendix \ref{Appendix:P}):

\begin{equation}\label{eq:PaL}
\begin{aligned}
\tilde{P}_a(x,s|x_0)&=\tilde{P}(x,s|x_0)-\tilde{P}(x,s|-x_0)=\\
&=\tilde{P}_{\rm nc}(x,s|x_0)+\frac{\tilde{P}_{\rm res}(s|x_0)}{1-\tilde{P}_{\rm res}(s|x_r)}\tilde{P}_{\rm nc}(x,s|x_r)-\tilde{P}_{\rm nc}(x,s|-x_0)-\frac{\tilde{P}_{\rm res}(s|-x_0)}{1-\tilde{P}_{\rm res}(s|x_r)}\tilde{P}_{\rm nc}(x,s|x_r)\\
&=\tilde{P}_{\rm nc}(x,s|x_0)-\tilde{P}_{\rm nc}(x,s|-x_0)
\end{aligned}
\end{equation}
where we exploit the fact that $P_{\rm res}(s|x_0)=P_{\rm res}(s|-x_0)$, this can be checked from Eqs. \eqref{eq:Pres1} and \eqref{eq:Pres2}. As expected, $P_a(x,t|x_0)$ depends only on the probability $P_{\rm nc}(x,t|x_0)$ of no resetting  or, equivalently, the probability 
density for particles to be evaporated with constant rate $r$ in the region $(0,a).$ Its Laplace transform is computed by plugging Eq. \eqref{eq:Pnc2} in Eq. \eqref{eq:PaL} and it reads

\begin{equation}\label{eq:Pa1}
\begin{aligned}
&\tilde{P}_{a}(x,s|x_0)=\left[D\left(\mu\sinh\left(a\nu\right)+\nu\cosh\left(a\nu\right)\right)\right]^{-1}\\
&\begin{cases}
\sinh(x\nu)\left[\nu\cosh\left(\nu(a-x_0)\right)+\mu\sinh\left(\nu(a-x_0)\right)\right]/\nu &\mbox{for}\,\,x\in [0,x_0),\\
\sinh(x_0\nu)\left[\nu\cosh\left(\nu(a-x)\right)+\mu\sinh\left(\nu(a-x)\right)\right]/\nu &\mbox{for}\,\,x\in [x_0,a),\\
e^{\mu(a-x)}\sinh(x_0\nu)&\mbox{for}\,\,x\ge a.
\end{cases}
\end{aligned}
\end{equation}
Because of absorption at $x_r$ the density of particles is not conserved. The fraction of non-absorbed particles is described by the survival probability ${S(t|x_0)=\int_0^\infty \mathrm{d}x\,P_a(x,t|x_0)}$
whose Laplace transform, computed integrating Eq. \eqref{eq:PaL}, is given by \begin{equation}\label{eq:SL}
\tilde{S}(s|x_0)=\frac{1}{D\nu^2}\left[1-\frac{\nu\cosh\left(\nu(a-x_0)\right)+\mu\sinh\left(\nu(a-x_0)\right)-r\sinh(x_0\nu)/D\mu}{\mu\sinh\left(a\nu\right)+\nu\cosh\left(a\nu\right)}\right].
\end{equation}
Figure \ref{Fig:Survival} shows the comparison between simulations for $S(t|x_0)$ (solid line) and numerical inverse Laplace transform of Eq. \eqref{eq:SL} (symbols).
It can be easily checked that, in the case $r=0$ corresponding to the presence of the sole absorption at the boundary $x_a=0$, ${\tilde{S}(s|x_0)=(1-\exp(-x_0\mu))/(D\mu^2)}$, whose inverse Laplace transform ${S(t|x_0)=\mathrm{erfc}\left(x_0/\sqrt{4Dt}\right)}$ reproduces a well-known result, see, e.g., Ref. \cite{redner_2001}.

Another quantity of physical interest is the first-passage probability to the origin at time $t$, indicated here by $F(t|x_0)$. By definition, this quantity is actually related to $S$ via

\begin{equation}
S(t|x_0)=1-\int_0^t\mathrm{d}\tau\,F(\tau|x_0),
\end{equation}
from which, in terms of the Laplace transform, becomes ${\tilde{F}(s|x_0)=1-s\,\tilde{S}(s|x_0)}$. Equivalently, $F$ may be computed also exploiting the fact that the first-passage of a particle to the origin corresponds to its absorption. Therefore, the first passage probability can be expressed as the sum of two contributions: the flux of particles to the absorbing wall and the flux of particles that get reset, see Ref. \citep{redner_2001}, namely

\begin{equation}\label{eq:F1}
\begin{aligned}
\tilde{F}(s|x_0)&=D\,\frac{\partial \tilde{P}_a(x,s|x_0)}{\partial x}\Bigg|_{x=0}+\int_0^\infty \mathrm{d}x\,r_c(x)\,\tilde{P}_a(x,s|x_0)\\
&=\frac{\mu\sinh\left((a-x_0)\nu\right)+\nu\cosh\left((a-x_0)\nu\right)}{\mu\sinh\left(a\nu\right)+\nu\cosh\left(a\nu\right)}\,+\frac{r}{D}\frac{\sinh\left(x_0\nu\right)}{\mu\nu^2}\frac{\nu\mu\sinh\left((a-x_0)\nu\right)+\mu^2\cosh\left((a-x_0)\nu\right)+r/D}{\mu\sinh\left(a\nu\right)+\nu\cosh\left(a\nu\right)}.
\end{aligned}
\end{equation}
Because of absorption, the aforementioned currents are only transient and vanish at long times.
In the absence of resetting $r=0$,  Eq. \eqref{eq:F1} yields  $\tilde{F}(s|x_0)=e^{-x_0\sqrt{s/D}}$ which is the Laplace transform of the inverse Gaussian $F(t|x_0)=x_0(4\pi D t^3)^{-\frac{1}{2}}e^{-x_0^2/4Dt}$, see Ref. \cite{redner_2001}.

Using the above exact results, we compute the fraction $Q_{\rm res}(x,t|x_0)$ of particles starting from $x_0$ that gets absorbed because of resetting from a certain position $x$  (with $0<x\le a$) within the time interval $(0,t)$, i.e., 
\begin{equation}
\begin{aligned}
Q_{\rm res}(x,t|x_0)&=r_c(x)\int_0^t\mathrm{d}\tau\, P_a(x,\tau|x_0)\\
&=r\,\theta(a-x)\int_0^t\mathrm{d}\tau\, P_a(x,\tau|x_0);
\end{aligned}
\end{equation}
the Laplace transform $\tilde{Q}_{\rm res}(x,s|x_0)$ of this quantity with respect to time $t$, using the expression of $\tilde{P}_a(x,s|x_0)$ in Eq. \eqref{eq:Pa1}, reads

\begin{equation}
\begin{aligned}
\tilde{Q}_{\rm res}(x,s|x_0)&=r_c(x)\,\frac{\tilde{P}_a(x,s|x_0)}{s}\\
&=\frac{r\sinh\left(x_<\nu\right)}{s\,D\nu}\frac{\nu\cosh\left((a-x_>)\nu\right)+\mu\sinh\left((a-x_>)\nu\right)}{\mu\sinh\left(a\nu\right)+\nu\cosh\left(a\nu\right)}
\end{aligned}
\end{equation}
with $x\in(0,a)$, $x_>=\max(x,x_0)$ and $x_<=\min(x,x_0)$. Unlike the quantities $P_a(x,t|x_0)$, $S(t|x_0)$ and $F(t|x_0)$,  $Q_{\rm res}(x,t|x_0)$ does have a well defined stationary limit, which is given by
\begin{equation}\label{eq:rho_inf}
\begin{aligned}
Q_{\rm res}(x|x_0)&=\lim_{t\rightarrow\infty}Q_{\rm res}(x,t|x_0)\\
&=\lim_{s\rightarrow 0}s\,\tilde{Q}_{\rm res}(x,s|x_0)\\
&=\sqrt{\frac{r}{D}}\frac{\cosh\left((a-x_>)\sqrt{\frac{r}{D}}\right)\sinh\left(x_<\sqrt{\frac{r}{D}}\right)}{\cosh\left(a\sqrt{\frac{r}{D}}\right)},
\end{aligned}
\end{equation}
where we denote the stationary distribution with $Q_{\rm res}(x|x_0)$, by dropping the $t$ dependence in $Q_{\rm res}(x,t|x_0)$.
The existence of a stationary limit of $Q_{\rm res}(x,t|x_0)$ derives from the fact that we consider only trajectories of resetting particles.
Finally, the fraction of particles that get absorbed by resetting is expressed according to 
\begin{equation}
\eta_{\rm res}(y_0=x_0/a)\equiv\int_0^a\mathrm{d}x\,Q_{\rm res}(x|x_0).
\end{equation}
Hence, by integrating Eq. \eqref{eq:rho_inf}, we obtain
\begin{equation}\label{eq:eta}
\eta_{\rm res}(y_0)=1-\frac{\cosh((1-y_0)\rho)}{\cosh\rho}
\end{equation} 
with $y_0=x_0/a$. In the context of RNA polymerase, $\eta_{\rm res}$ measures the efficiency of the polymerase during RNA backtracking through cleavage. As expected, $\eta_{\rm res}(y_0)$ is a monotonically increasing function of $y_0$, this is due to the fact that the larger is $y_0$, more time particles will spend in the absorbing region: the maximum is attained at $\eta_{\rm res}(y_0=1)=1-(\cosh\rho)^{-1}<1$ and its infimum is $\eta_{\rm res}(y_0=0)=0$ corresponding to the physically irrelevant case of total absorption at $t=0$. In the limit of large absorption, $\eta_{\rm res}\approx 1-e^{-y_0\rho}$, so that $\eta_{\rm res}\approx 1$ for values $y_0 \simeq \rho^{-1}$, meaning that absorption is mostly due to resetting (evaporation). In particular, if one considers the limiting case $a\rightarrow\infty$, the distribution reduces to $\tilde{P}_a(x,s|x_0)=\sinh(x_<\nu)e^{-x_>\nu}/D\nu$, whose inverse Laplace transform is the expected $P_a(x,t|x_0)=e^{-rt}(4Dt\pi)^{-\frac{1}{2}}\left(\exp\left[-(x-x_0)^2/4Dt\right]-\exp\left[-(x+x_0)^2/4Dt\right]\right).$ It then follows that $S(t|x_0)=e^{-rt}\erf\left(\frac{x_0}{\sqrt{4Dt}}\right)$ and ${\eta_{\rm res}(y_0)=1-e^{-y_0\rho}}$; as we report below, this case corresponds to the polymerase Pol II TFIIS.

The analysis presented above focused on the case $x_0 \in (0,a)$, but it can be readily generalized for $x_0>a$, with the following results: the survival probability (instead of  Eq. \eqref{eq:SL}) reads
\begin{equation}
\tilde{S}(s|x_0)=\frac{1}{s}\left[1-\frac{e^{\mu(a-x_0)}(r\cosh(a\nu)+s)}{D\nu(\mu\sinh\left(a\nu\right)+\nu\cosh\left(a\nu\right))}\right].
\end{equation}
Hence, we compute $Q_{\rm res}(x,s|x_0)$ as
\begin{equation}
\tilde{Q}_{\rm res}(x,s|x_0)=\frac{r\,e^{\mu (x_0-x)}}{D\,s}\frac{\sinh(x\nu)}{\mu\sinh\left(a\nu\right)+\nu\cosh\left(a\nu\right)}
\end{equation}
with $x\in(0,a)$ and its stationary limit reads

\begin{equation}
Q(x|x_0)=\sqrt{\frac{r}{D}}\frac{\sinh\left(x\sqrt{\frac{r}{D}}\right)}{\cosh\left(a\sqrt{\frac{r}{D}}\right)}.
\end{equation}
As above, by integrating $Q_{\rm res}(x|x_0)$ over $x\in(0,a)$ we get the fraction of particles that gets absorbed because of resetting at all times, i.e.,

\begin{equation}\label{eq:eta2}
\eta_{\rm res}(y_0)=1-\frac{1}{\cosh(\rho)}
\end{equation}
does not depend on $y_0$. 
Because of recurrence, the Brownian motion will eventually hit the resetting region, independently of its initial position. 
Collecting the above result in Eqs. \eqref{eq:eta} and \eqref{eq:eta2} we derive a general expression for $\eta_{\rm res}(y_0)$ in Eq. \eqref{eq:reseff2} for any $y_0>0$.


Finally, we report in Table~\ref{tab:1} the parameter values used in Fig. \ref{fig:Pol} in the main text for the analytical predictions and numerical simulations of RNA polymerases  Pol I, Pol II, and Pol II TFIIS.

\begin{center}
\begin{table}
  \begin{tabular}{ | c | c | c | c |}
    \hline
    Type & $a\,\,(\text{nt})$ & $r\,\,(1/\text{s})$ & $D\,\,(\text{nt}^2/\text{s})$ \\ \hline
    Pol I & 20 & 0.02 & 0.21\\ \hline
    Pol II & 10 & 0.01 & 0.54 \\ \hline
    Pol II TFIIS& $\infty$ & 0.076 & 1.6\\ \hline
  \end{tabular}
\caption{Parameters associated to the different species of polymerase, see Ref. \citep{lisica2016mechanisms}, corresponding to Fig. \ref{fig:Pol} c-d. The length unit is a nucleotide, $1$ nt =$0.3$ nm.}
\label{tab:1}
\end{table}
\end{center}
\newpage
\twocolumngrid

%


\end{document}